# T-DB: Toward Fully Functional Transparent Encrypted Databases in DBaaS Framework (Full Version)

Xiaofei Wang
National Computer Network Intrusion Protection Center, University of Chinese Academy of Sciences
Beijing, P.R. China
wangxiaofei14@mails.ucas.ac.cn

Qianhong Wu
School of Electronic and Information Engineering
Beihang University
Beijing, P.R. China
qianhong.wu@buaa.edu.cn

Yuqing Zhang*
National Computer Network Intrusion Protection Center, University of Chinese Academy of Sciences
Beijing, P.R. China
zhangyq@ucas.ac.cn

## ABSTRACT

Individuals and organizations tend to migrate their data to clouds, especially in a DataBase as a Service (DBaaS) pattern. The major obstacle is the conflict between secrecy and utilization of the relational database to be outsourced. We address this obstacle with a Transparent DataBase (T-DB) system strictly following the unmodified DBaaS framework. A database owner outsources an encrypted database to a cloud platform, needing only to store the secret keys for encryption and an empty table header for the database; the database users can make almost all types of queries on the encrypted database as usual; and the cloud can process ciphertext queries as if the database were not encrypted. Experimentations in realistic cloud environments demonstrate that T-DB has perfect query answer precision and outstanding performance.

## 1. INTRODUCTION

Cloud computing is an Internet-based commercial computing pattern based on networking, virtualization and distributed technology, where shared resources are allocated on demand [1, 2]. In such a dynamic and open environment, each cloud user has easy access to high computation speed and large storage capacity. Gradually, cloud computing is widely accepted by industry around the world. According to the survey data announced by International Data Corporation (IDC) [3] in early 2016, worldwide spending on public cloud services will grow at a 19.4% compound annual growth rate from nearly 70 billion dollars in 2015 to more than 141 billion dollars in 2019. Among them, cloud IT infrastructure spending will reach 53.1 billion dollars, which will be 46% of total expenditures on enterprise IT infrastructure. In the same year, a latest report from the famous American IT market research agency TBR (Business Research Technology) [4] predicts that worldwide public cloud revenue will increase from 80 billion dollars in 2015 to 167 billion dollars in 2020. We can see that a notable growth space still exists in the highly competitive public cloud market. Various cloud-based products are emerging in endlessly. Data outsourcing is one of its most important applications as well as a research focus of the trusted cloud computing.

In recent years, more and more individuals and organizations have been migrating their private data into a Cloud Service Provider (CSP). About data outsourcing, the DataBase as a Service (DBaaS) [5, 6, 7] has become one of the most representative service patterns, e.g., Amazon RDS, Google Cloud SQL, Microsoft Windows Azure SQL Database, Tencent CDB, etc. They all are committed to a cloud database platform similar to the well-known Relational DataBase Management System (RDBMS).

DBaaS conveys a great number of conveniences and benefits, but once stored on the cloud storage servers, the outsourced private data (e.g., personal health care records, financial transaction, etc.) could be disclosed or abused. The security issues of the outsourced databases are increasingly important. The open challenge is to simultaneously achieve database secrecy and full utilization.

Currently, in most public clouds, the approach for the CSP to protect cloud database security is by enforcing cloud-centralized security strategies including resource isolation, authorized access and cloud encryption, etc. For some commercial products, such as SecureDB or Windows Azure SQL Database, encryption is considered critical to prevent various malicious intrusions from external attackers.

For the cloud-centralized security strategies to work, it is necessary that the CSP is willing and able to secure the outsourced database, which is a quite strong assumption in practice. These cloud-centralized security strategies cannot prevent the malicious CSPs from exploiting the outsourced data for commercial benefits. Also, they fail to prevent malicious inner attackers, e.g., the corrupted employees. Furthermore, if the CSP is hacked, then the outsourced database could be fully exposed to attackers.

One possible solution is for the owner to encrypt her database so the CSP is no longer involved in secret key-dependent operations. The database is encrypted before being outsourced to the CSP. This, however, incurs a utilization problem for the encrypted database and requires elegantly designed encryption schemes. So far, encryption technologies have been developed to support Equality Query [8, 9, 10], Range Query [11, 12, 13], Aggregation Query [14, 15, 16] and Fuzzy Query [17, 18, 19]. Yet, none of them could support the queries as the plaintext database does. Fully homomorphic encryption techniques [20] theoretically permit arbitrary operations over encrypted data, but the computational complexity of encryption/decryption is prohibitively high [21].

In 2011, a research team [22] from MIT released the famous CryptDB. It supports most manipulations on encrypted databases based on Structured Query Language (SQL). However, CryptDB is not available in the DBaaS framework. Specialized encryption algorithms are adopted respectively to interpret various kinds of SQL statements. The SQL operations cannot be performed directly. Instead, the database owner sends the secret keys to the cloud servers for the onion decryptions when processing a query, which introduces security risks similar to those in cloud-centralized security strategies. Variants [23, 24] using different encryption algorithms lead to serious performance degradation and heavy reliance on the trusted proxies. Moreover, they require changing the access mode of database users as well as the structure of the cloud database. Other variants (e.g., [25, 26]) simply allow limited types of ciphertext manipulations.

*Corresponding author.

In view of the limitation in performance of homomorphic techniques and the incompatibility in functionality of existing solutions, our T-DB in this paper is a new comprehensive encrypted database system with database-owner-centralized security strategies under the standard DBaaS framework. It is the first time that all the known types of potential ciphertext queries (i.e., equality, range, aggregation and fuzzy queries) are truly covered through a single cipher, and our average efficiency of ciphertext queries is actually better than that of homomorphic encryptions. The contributions of this paper are summarized as follows:

1) We strictly follow the DBaaS framework and present T-DB equipped with distinguishing features. It can perform almost all of data manipulations over outsourced encrypted databases. T-DB is built for the database users who make queries on plaintext data; the database owner who outsources the encrypted database to the CSP, translates and forwards queries from the database users to the CSP, and decrypts and returns answers to the users; and the CSP who executes encrypted queries on encrypted databases as if they were not encrypted. All the operations involving secret keys are performed by the database owner, who can hence be assured that the database is under her control. The database users and the CSP just work as usual and there is no need for a database user to change the way he accesses the database. Thus T-DB is implementation-friendly and user-friendly.

2) With the help of User-Defined Function (UDF), we deployed T-DB on an unmodified cloud RDBMS server within 30 lines of codes at the cloud server side and a local client machine using standard SQL syntax at the database owner side. We demonstrate that T-DB is secure under two threat models. This is achieved on the premise of supporting almost all types of queries and dynamic updates over an encrypted database.

3) A novel order preserving encryption algorithm with additivity and an SQL-based syntax interpreter are constructed, which allow direct comparison and addition to be executed transparently on ciphertexts by one-time encryption and statement translation. Formal security analyses indicate that this cipher tool can resist weak chosen-plaintext attack.

4) Experimental results show that T-DB supports more queries and achieves a precision rate (or a recall rate) of 100%. Compared with CryptDB under the same precision, query processing in T-DB is faster and its decryption time for query results is shorter. Also, the T-DB applicability is validated on Microsoft Windows Azure SQL Database.

The remainder of this paper is organized as follows:

Firstly, Section 2 and Section 3 give an overview of the model and architecture. Then in Section 4 and Section 5, two core modules are proposed. Performance evaluations are provided in Section 6. Finally, we fully discuss the system security, some special cases, related works and future direction in Section 7, and conclude the study in Section 8.

# 2. BASIC MODEL

In this section, we define clearly the generalized system model of DBaaS framework and its major security threats, requirements and design objectives. The database owner should emphatically consider two problems during cloud outsourcing. For one thing, how to outsource a database to the cloud, in order to ensure the data storage confidentiality. For another, after the database is outsourced to the cloud, how database users normally get access to their data.

## 2.1 System Model

Data outsourcing is to outsource complex data structure, data storage and data computation to the CSP. The major objective of outsourcing data is to allow the cloud users avoid the lack of local resources, and minimize the cost of deployment, operation and maintenance in data application systems. As one of the most important patterns, database resources are pooled by DBaaS and delivered to database users through network in the form of cloud service. The cloud is responsible for the storage of structured business data, and provides comprehensive database functionalities, including data analysis and data processing.

As an enterprise-level cloud storage service, DBaaS framework allows a database owner inside an enterprise to outsource her internal database to the CSP, thus reducing the cost of management and maintenance in the enterprise. The generalized system model of DBaaS is shown in Figure 1, which is composed of three types of entities, i.e., Database User, Database Owner and CSP. More details are as follows:

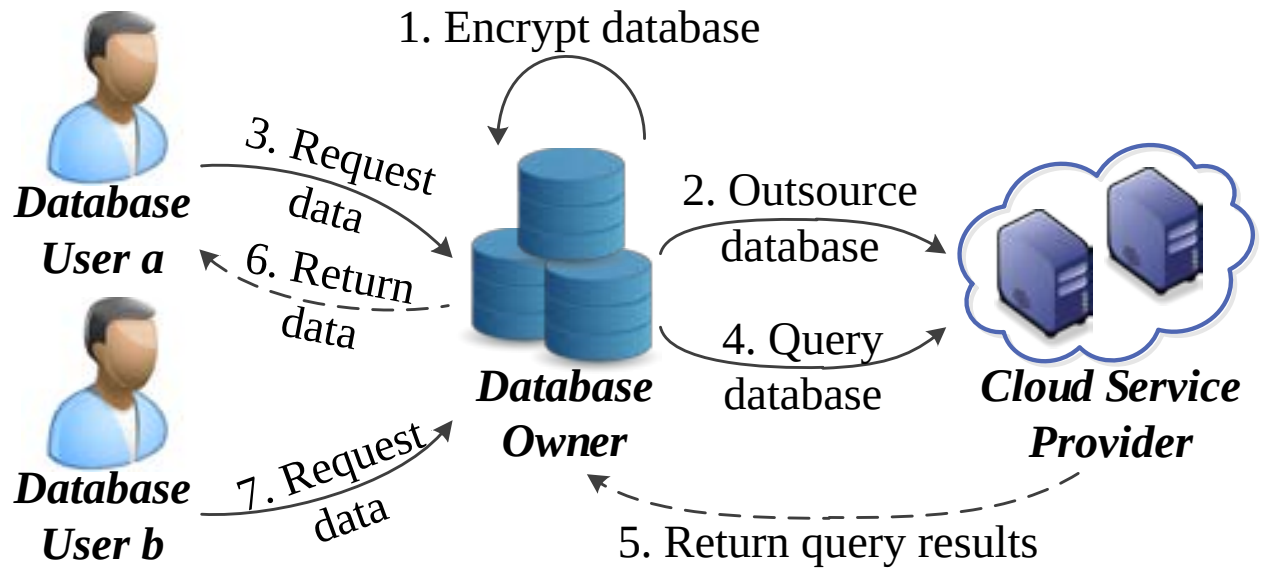

**Figure 1: System model.** Solid arrows indicate the request sent to the CSP and dotted arrows represent the reply returned by it.

1) The database user is the requestor of data records. Different database users have different access permissions. At any time, only the authorized subset can be accessed using a valid login password.

2) The database owner is the provider of original plaintext databases. She outsources the databases to the CSP after encryption and requests queries on the behalf of the database users. For a multi-user application scenario, the database owner takes charge of dividing user groups and assigning role permissions. Also, the database owner may enforce strict access control policies according to the enterprise's security requirements. For instance, after a database user (e.g., user *a*) successfully logs into the system using his password $P_a$, the database owner will perform database queries within the authorized scope of the database user *a* and return the final query results to him.

3) The CSP is the service provider of cloud database storage. It is established on an unmodified RDBMS. Generally speaking, the CSP is untrusted and can be assumed as honest-but-curious. That is, although the CSP honestly follows the instructions of the system, it may be also interested in the database owner's private data for commercial gains.

## 2.2 Threat Model

At present, the information disclosure of traditional DBaaS is mainly shown as two trends: (i) The way to access outsourced data through network provides an opportunity for external attackers to compromise the CSP. The hackers can eavesdrop the private information of the outsourced database by exploiting security vulnerabilities in the B/S applications. (ii) Essentially, from the view of the CSP, the data outsourced to the cloud is stored as plaintexts. Since the sensitive information is exposed directly to the internal attackers in the CSP, data loss or data leakage often occurs. Even though the transmission process of data outsourcing is secure, the database owners lost physical control of their data.

Outsourced data can be encrypted by the CSP before storage. Strict isolation strategies and access control policies are also developed. In this way, the system may prevent the data leakage caused by the plaintext storage or the possible external attacks that break security boundary. But in this process, for one thing, the access mode of cloud users has been changed. Database users have to log into the cloud servers to access their data. For another, the CSP has to be trusted in the management of encryption keys and access permissions. Thus, to protect the secrecy of outsourced data, the database owner should encrypt the database before outsourcing. And to make database users normally access outsourced data, the encrypted database should support almost all types of common query manipulations.

As for the outsourcing flow of DBaaS in Figure 1, since the encryption and the decryption in outsourced databases are operated by the database owner, we assume that sufficient access control or other security mechanisms are applied at the database owner side, to make sure the secret keys will not be leaked. The system model of T-DB addresses two types of possible security threats:

**Threat model 1.** The cloud service providers furtively steal the private data of the database owner. And the case that a third-party attacker intrudes the CSP to compromise data secrecy is also covered in this model. This kind of attacker can seize encrypted data and the database structure from the CSP. And he can launch brute force attacks (i.e., ciphertext-only attacks) with this information.

**Threat model 2.** The database users collude with cloud service providers. This kind of attacker has the ability to control the query manipulations on the encrypted database. By constructing an appropriate SQL statement, more confidential information might be acquired, including specific query requests and partial query results. In this case, numerous pairs of plaintext-ciphertext can be analyzed. The attacker can launch a chosen-plaintext attack or a chosen-ciphertext attack, in order to guess encryption keys.

In the practical applications, the main source of attacks is the first type threat. On one hand, the attacker is easy to have access rights of the encrypted data stored in the cloud, but he is difficult to determine the corresponding relationship between plaintexts and ciphertexts. So by means of the brute force attack, it is pretty hard to guess the encryption keys. On the other hand, through a comprehensive consideration of the database structure and the results from multiple queries, the attacker can gradually infer an approximate distribution of the encrypted database. Then several statistical information about data items may be exposed. Hence, the security against the ciphertext-only attack is the fundamental security requirement.

T-DB is designed to protect the secrecy and utilization of outsourced data under the above two types of security threats. It does not ensure the data integrity, freshness, or the completeness of query results. And the attack on database owner/users' local machines is also outside of the scope of our paper. In addition, T-DB should be user-friendly and implementation-friendly. It is designed to be deployed on an unmodified CSP, and without changing the access mode of the database users, it requires that almost all types of data manipulations can be executed directly over the outsourced encrypted database.

# 3. SYSTEM OVERVIEW

The secure outsourcing in our T-DB works by executing various SQL-based data manipulations directly over encrypted databases. Its overall architecture is based on the DBaaS framework, consisting of three layers, as shown in Figure 2. Different from the trusted proxy in CryptDB, the database owner in T-DB architecture does not change the database structure, so no extra expense happens on cloud storage.

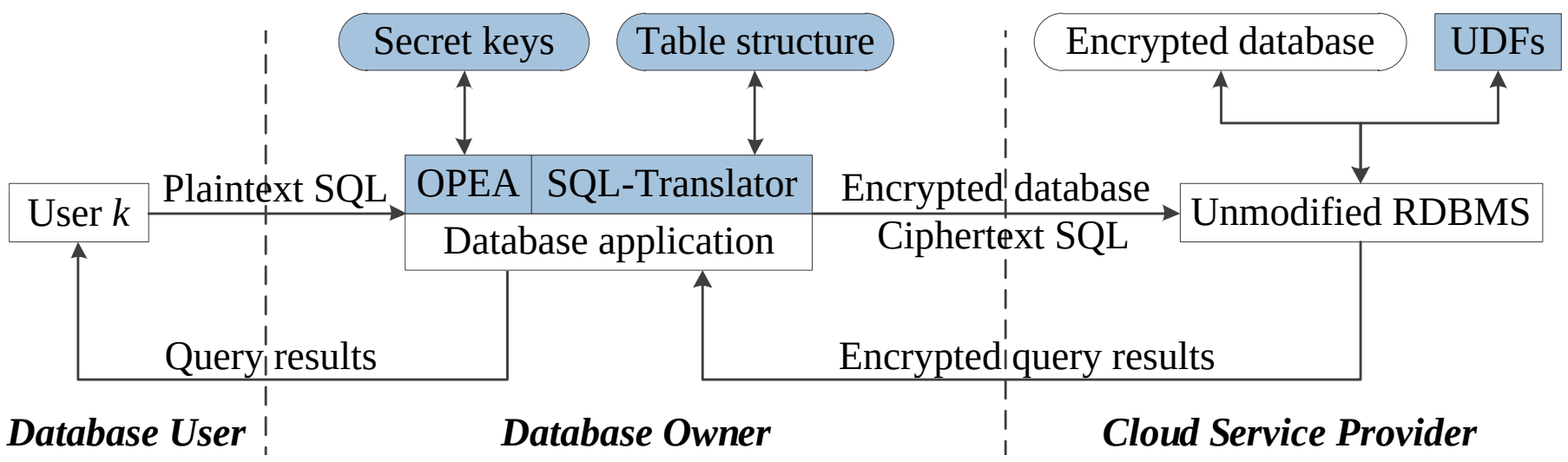

**Figure 2: Overall architecture of T-DB.** Rectangular and rounded boxes represent processes and data, respectively. Blue shading indicates components added by T-DB system. Arrows indicate the data transmission and the call relation in the process of outsourcing. Dashed lines are used to separate the scope of the database user, the database owner and the CSP.

## 3.1 Data Flow

The outsourcing flow for plaintext databases is as follows:

1) The database owner secretly generates and keeps the secret keys. A database is encrypted before it is outsourced. After that, no copy of the plaintext database and the encrypted database is preserved by the database owner.

2) The CSP receives and stores the encrypted database. Query-related UDFs are also built and optimized in advance.

The data manipulations on encrypted databases are as follows:

1) The database user $k$ uses his password $P_k$ to log into the system, and submits a plaintext SQL to the database owner.

2) According to access control policies, the database owner checks whether the query request is authorized and from a legal user. If it is an unauthorized access, then this manipulation is refused and a security alert is returned.

3) The database owner translates the plaintext SQL statement into ciphertext SQL statements, and sends them to the CSP.

4) Appropriate UDFs are invoked by the CSP to execute the ciphertext SQL over the encrypted database, and several encrypted query results are returned.

5) The database owner receives and decrypts the result set, and returns the plaintext query results to the database user $k$.

## 3.2 Module Design

Figure 2 also shows the data processes in T-DB. Compared with existing models for outsourcing secured data, a distinguishing feature of T-DB is that all crypto-operations are performed by the armed database owner so that they are transparent to both the database users and the CSP. More specifically, the database owner fulfills the following four tasks:

1) The database owner keeps the table structure of the original plaintext database, such as the database diagram, field type, field length, scope of data items, and so on.

2) In the pre-processing stage, the database owner adopts two cipher tools to encrypt and outsource the plaintext database. The data items are encrypted with an SQL-oriented encryption algorithm, and the table names and column names are encrypted with a symmetric encryption algorithm or a collision-resistant salted hash function. The database owner always encrypts different data items with different secret keys. These keys are assigned through a pseudo-random generator with a seed key to simplify key management.

3) In the manipulation stage, the database owner translates the plaintext SQL statements by an interpreter (e.g., SQL-Translator) into a set of ciphertext SQL statements with the same semantics. For more details about translation methods, please see Section 5.

4) In the post-processing stage, the database owner decrypts the CSP's answer of the ciphertext SQL statements and returns the final results to the database user.

To ensure data secrecy and normal access to cloud storage, the database owner needs to deploy two additional modules, namely cipher module and translation module. The former is to achieve secrecy so that the database is accessible via its owner, and database updates are supported. The latter is to guarantee that neither the CSP nor database users are affected due to the cipher module.

# 4. CIPHER MODULE: OPEA

The cipher module executes the encryption when data are outsourced to the CSP, which ensures that the ciphertext manipulability is well preserved in DBaaS. This implies that the encryption algorithm is somewhat homomorphic. Observe that the majority of data manipulations in relational databases can always be converted into range queries and aggregation queries. Therefore, for T-DB system to work, its cipher module needs to simultaneously meet order-preservation and additivity. These requirements thus define a novel cipher tool of Order Preserving Encryption with Additivity (OPEA).

In this section, the basic idea of OPEA is introduced first, followed by its boundary generation sub-algorithm, encryption sub-algorithm and decryption sub-algorithm. Several properties are then formally described in detail. We also provide some simplified and extended versions. The end of the section further analyzes their security and limitation.

## 4.1 Basic Idea

An appropriate SQL-oriented cryptographic technique is needed to execute ciphertext manipulations over outsourced encrypted databases, and thus the cipher tool should be order preserving as well as additive. To this end, we introduce a new idea that is closely related to the order preserving encryption.

Order Preserving Encryption (OPE) is a cryptographic technique which can keep the plaintext order after encryption [11]. The first order preserving encoding function for numeric data was proposed by Agrawal et al. [11] in 2004, whose one-to-one mapping encoding is inefficient and vulnerable to the statistical attack. mOPE [12] works on an interactive encryption protocol to build an encrypted balanced search tree. It hides the order in its ciphertexts and uses additional operations to finish the comparison. Popa et al. [12] showed that mutable ciphertexts are necessary for the ideal IND-OCPA security [27]. Deterministic OPE algorithms always lead to the leakage of plaintext distributions, particularly for those asymmetric OPEs that can be broken easily. The attacker first encrypts every value in the plaintext domain with public keys. Then the relationship between a plaintext and its corresponding ciphertext is established. Any ciphertext can be matched quickly with binary search, thus its plaintext is deciphered successfully. One straightforward solution to the above drawback is to map each plaintext value to a random number in the irregular range with a certain probability, that is, the plaintext should be encrypted by the one-to-many mapping. For example, Reddy and Ramachandram designed a Randomized OPE (called ROPE [28]), which follows the mOPE in [12]. It leaks nothing except the comparison order. On the premise that the whole distribution of the plaintext domain is selected in advance, a recent LiuOPE [29] draws support from the nonlinear space split and some one-to-many mapping functions to extend the message space. It can resist ciphertext-only attacks or particular chosen-plaintext attacks. The one-to-many mapping in LiuOPE is obviously boundary-crossing. Precision loss and the leakage of boundaries of ciphertext partitions might appear in equality queries.

The comparison order among plaintext values is fully preserved by OPE, whereas its order-preservation is no longer valid after several ciphertexts are added together. Therefore, although most of the existing OPE schemes can be incorporated in equality queries, range queries or its relevant extensions, they cannot run any aggregation queries that contain algebraic operations without decryption, such as SUM function and AVG function. The application of OPE techniques in data manipulations of encrypted databases is limited due to the lack of additivity. The main reason is that the comparison and the aggregation in a practical query request are usually operated in an irregular way. In order to facilitate the composite queries, ciphertexts should keep in order after being added together. Besides, although a subtraction can offer the same query as a comparison, the RDBMS row-based access determines that the data items in the same column are hardly sorted through subtraction.

We illustrate an OPEA, inspired by well-established OPE algorithms [11, 12, 28, 29] with one-to-many mapping, that allows both comparison and summation to be operated on ciphertexts. Numerical order can still be preserved after several ciphertexts are added together. Specifically, the OPEA encryption function $E$: $X \rightarrow Y$ meets the following conditions:

1) Order-Preservation: $\forall a, b \in X$, $a<b$ ***iff*** $E(a)<E(b)$.

2) Additive order-preservation: $\forall a, b, c \in X$, ***if*** $a+b<c$ ***then*** $E(a)+E(b)<E(c)$.

where $X$ and $Y$ represent the plaintext and the ciphertext domain respectively. The algorithmic design of OPEA is relatively straightforward, but technically it is quite suitable for the encrypted database system to develop.

The mapping structure of the OPEA encryption is illustrated in Figure 3. A discrete ciphertext domain is composed of sequential partitions, and there is a non-empty mutable interval between two adjacent partitions. As shown in this figure, for the one-to-many encryption, a plaintext value $b$ is mapped to a random number in the partition $[L_b, U_b]$, where $L_b$ and $U_b$ are the lower boundary and the upper boundary of this partition respectively. Intervals ensure that no overlap exists between any two partitions, that is, for any value $c$ ($c>b$), its lower boundary $L_c$ should be strictly greater than the upper boundary $U_b$. Similarly, for any value $a$ ($a<b$), its upper boundary $U_a$ should be strictly less than the lower boundary $L_b$. Meanwhile, in order to meet the additive order-preservation, the inequation $U_a+U_b<L_c$ should strictly hold.

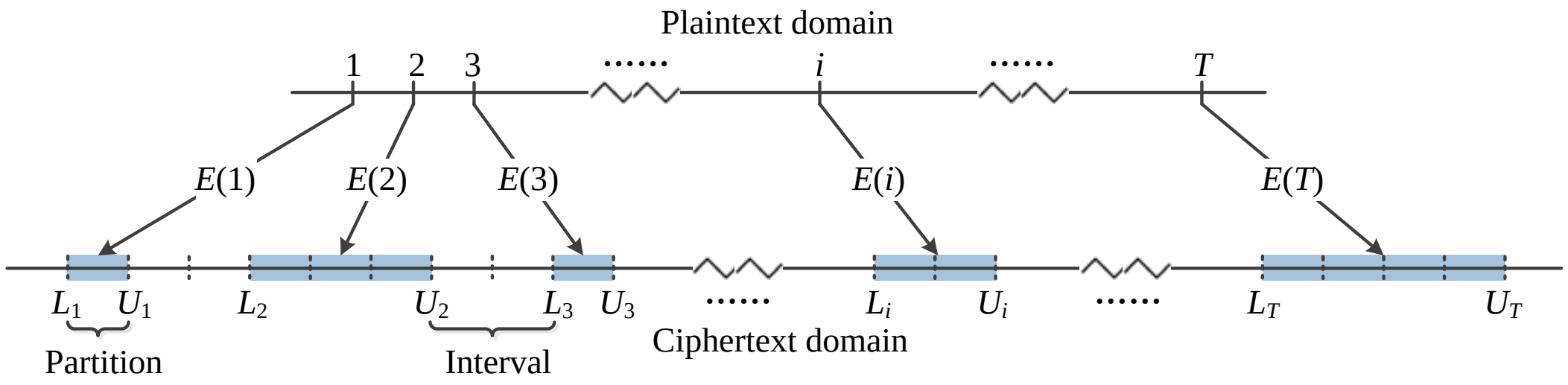

**Figure 3: One-to-many mapping in OPEA encryption.**

Throughout the paper, we assume that the lower (resp. the upper) boundary of the $i$th partition can be written as $L_i$ or $L[i]$ (resp. $U_i$ or $U[i]$) for easy presentation.

## 4.2 Algorithm Description

Without loss of generality, we define the plaintext domain of OPEA as a part of positive integers, that is, $X \subseteq N^+$. According to the basic ideas in Section 4.1, OPEA can be summarized formally as a symmetric encryption algorithm, i.e., (**BoundaryGen**, **Enc**, **Dec**), including three sub-algorithms:

1) Boundary generation sub-algorithm (Algorithm 1).

This sub-algorithm takes $\boldsymbol{R}$ and $\sigma$ which are selected randomly as system keys to return two finite integer sets ($\boldsymbol{L}$ and $\boldsymbol{U}$). $\boldsymbol{R}=\{R_i\}$ ($1 \le i \le T$) is a set of non-negative integers, and $\sigma$ is a positive integer with $\sigma > \max_{1 \le i \le T}\{R_i\}-R_1$, where $T$ is the maximum value in the plaintext domain. $\boldsymbol{L}$ (resp. $\boldsymbol{U}$) contains the lower boundaries (resp. the upper boundaries) of every partition in the ciphertext domain.

**Algorithm 1** ($\boldsymbol{L}$, $\boldsymbol{U}$)$\leftarrow$**BoundaryGen**($\boldsymbol{R}$, $\sigma$)

Set $L[1]=\sigma$, $U[1]=L[1]+R_1$
**for** $t$=2 to $T$ **do**
  Compute $L[t]=\max_{1 \le i < t}\{U[i]+U[t-i]\}$
  Compute $U[t]=L[t]+R_t$
**end for**
**return** $\boldsymbol{L}$, $\boldsymbol{U}$

The first partition $[L_1, U_1]$ is selected randomly by the database owner in its initialization phase, and the lower and upper boundaries of subsequent partitions are computed iteratively. Finally, the boundary sets generated by this sub-algorithm divide the ciphertext domain into some discrete partitions. The length of the partition $[L_i, U_i]$ for the plaintext $i$ is determined by the random number $R_i=U_i-L_i$, and the introduced interval between two sequential partitions can ensure non-overlap and additive cipher. We clarify the above mapping relation in Figure 3, where $R_1=R_3=1$ and $R_2=3$.

2) Encryption sub-algorithm (Algorithm 2).

This quasi-probabilistic sub-algorithm takes a plaintext value $m$ and two boundary sets ($\boldsymbol{L}$ and $\boldsymbol{U}$) to return its ciphertext $c=E(m)$. The function RandomSelect(·) is defined to select a random element from the parameter set $\{L[m], L[m]+1, \ldots, U[m]\}$ and return it as ciphertext. Intuitively, the encryption function keeps the numerical order of plaintexts, which will be fully discussed in Section 4.3.

**Algorithm 2** $E(m)\leftarrow$**Enc**$(m, \boldsymbol{L}, \boldsymbol{U})$

```
Set E(m)=RandomSelect(L[m], L[m]+1, L[m]+2, …, U[m])
return E(m)
```

3) Decryption sub-algorithm (Algorithm 3).

This deterministic sub-algorithm takes the boundary set $\boldsymbol{L}$ and a ciphertext value $c$ to return its plaintext $m=D(c)$. The ciphertext is decrypted by means of a reverse mapping and the process is similar to binary search method.

**Algorithm 3** $D(c)\leftarrow$**Dec**$(c, \boldsymbol{L})$

```
Set left=1
Set right=T
while left<right do
  Set mid=(left+right)/2
  if L[mid]≤c<L[mid+1] then
    Set D(c)=mid
    return D(c)
  else if c<L[mid] then
    Set right=mid-1
  else
    Set left=mid+1
  end if
end while
Set D(c)=left
return D(c)
```

## 4.3 Algorithm Property

Some properties about correctness, order-preservation and additivity of the OPEA algorithm are given below.

1) Correctness. As an order preserving encryption algorithm, the encryption function and the decryption function in OPEA output strictly the correct pairs of plaintext-ciphertext and always preserve their order. Since there is no overlap between different partitions and every ciphertext corresponds to a unique partition, the index of this ciphertext partition (i.e., the plaintext value) can be decrypted correctly by a lookup of the boundary set $\boldsymbol{L}$. Then correctness of OPEA is proven as follows:

$$\forall m\in X,\ m\leftarrow \mathbf{Dec}(\mathbf{Enc}(m, \boldsymbol{L}, \boldsymbol{U}), \boldsymbol{L})$$

2) Order-Preservation. Several lemmas are first established.

**Lemma 1 (Boundary effectiveness).** *For any integer t* ($t$=1, 2, …, $T$), *its partition boundary satisfies that* $L[t]\leq U[t]$.

***Proof.*** From Line 4, Algorithm 1, let us consider that if $R_t$ is a non-negative integer, then

$$L[t]\leq U[t] \qquad \square$$

**Lemma 2 (Boundary monotonicity).** *For any integer t* ($t$=1, 2, …, $T$-1), *its partition boundary satisfies that* $L[t]<L[t+1]$ *and* $U[t]<U[t+1]$.

***Proof.*** Based on Lemma 1, a conclusion follows that

$$U[t+1]\geq L[t+1]\geq U[1]+U[t]>U[t]\geq L[t]$$

where the second derivation step is due to Line 3, Algorithm 1. So the boundary sets $\boldsymbol{L}$ and $\boldsymbol{U}$ are monotonic. □

Next, we attempt to show several theorems for OPEA.

**Theorem 1 (Order-Preservation).** *For any integer t* ($t$=1, 2, …, $T$-1), *the encryption function satisfies that* $E(t)<E(t+1)$.

***Proof.*** In light of Lemma 2 and Algorithm 2, two conclusions follow that

$$L[t]\leq E(t)\leq U[t],\ L[t+1]\leq E(t+1)\leq U[t+1]$$

Notice that

$$U[t]<L[t+1]$$

Then it is true that

$$E(t)<E(t+1)$$

So Theorem 1 follows, and vice versa. The encryption function in OPEA is monotonic, that is,

$$\forall a, b\in X,\ a<b \textbf{ iff } E(a)<E(b) \qquad \square$$

**Theorem 2 (Additive order-preservation).** *For any integers* $a+b<c$ ($a$, $b$, $c\in X$), *the encryption function satisfies that* $E(a)+E(b)<E(c)$.

***Proof.*** Firstly, the following property holds

$$E(a)+E(b)\leq U[a]+U[b]\leq L[a+b]$$

where the second inequation is due to Line 3, Algorithm 1, i.e., $L[t]=\max_{1\leq i<t}\{U[i]+U[t-i]\}$.

And since

$$L[a+b]<L[c],\ L[c]\le E(c)$$

Then it is true that

$$E(a)+E(b)<E(c)$$

So Theorem 2 follows, and the additive order-preservation of OPEA is proven. □

## 4.4 Simplified and Extended Version

There are simplified and extended versions of the OPEA algorithm, which are discussed below.

### *4.4.1 Simplified Version of OPEA*

The one-time **BoundaryGen** needs $O(T^2)$ computations to generate $T$ pairs of boundaries. In order to reduce the local computational cost, the secret keys $R_i$ ($1\le i\le T$) should be generated in a non-decreasing order. A simplified boundary generation sub-algorithm $(\boldsymbol{L}, \boldsymbol{U})\leftarrow$ **SBoundaryGen**$(\boldsymbol{R}, \sigma)$ is developed as Algorithm 4.

**Algorithm 4** $(\boldsymbol{L}, \boldsymbol{U})\leftarrow$**SBoundaryGen**$(\boldsymbol{R}, \sigma)$

```
1: Set S=0
2: for t=1 to T do
3:   Compute L[t]=tσ+(t-1)R_1+S
4:   Compute U[t]=L[t]+R_t
5:   Compute S=S+R_t
6: end for
7: return L, U
```

From Line 3 to Line 5 in Algorithm 4, the lower and upper boundaries of ciphertext partitions can be calculated in linear time, which is proven as follows:

**Theorem 3 (Linear boundary function).** *For any integer t* ($1\le t\le T$)*, its lower and upper boundaries satisfy that*

$$L[t]=t\sigma+(t-1)R_1+\sum_{j=1}^{t-1}R_j \tag{1}$$

$$U[t]=L[t]+R_t \tag{2}$$

***Proof.*** The second mathematical induction is used to prove the correctness of Theorem 3.

Basis: In light of Line 1, Algorithm 1, Eq. (1) holds for $t$=1. $L[1]$ amounts to the equation

$$L[1]=1\cdot\sigma+(1-1)\cdot R_1=\sigma$$

Inductive step: Let us assume that Eq. (1) holds for any integer $t$ ($1<t\le k$, $k<T$), that is,

$$L[t]=t\sigma+(t-1)R_1+\sum_{j=1}^{t-1}R_j,\ 1<t\le k$$

Then for $t=k+1$, according to Line 3, Algorithm 1 and the definition of the lower boundary, $L[k+1]$ can be written as

$$L[k+1]=\max_{1\le i<k+1}\{U[i]+U[k-i+1]\}=\max\{U[1]+U[k],\ \max_{2\le i<k+1}\{U[i]+U[k-i+1]\}\}$$

where the left part can be deduced as

$$\begin{aligned}U[1]+U[k]&=L[1]+R_1+L[k]+R_k\\&=\sigma+R_1+k\sigma+(k-1)R_1+\sum_{j=1}^{k-1}R_j+R_k\\&=(k+1)\sigma+kR_1+\sum_{j=1}^{k}R_j\end{aligned}$$

And for any integer $i$ ($2\le i\le k$), the right part can be deduced as

$$\begin{aligned}&U[i]+U[k-i+1]\\&=L[i]+R_i+L[k-i+1]+R_{k-i+1}\\&=i\sigma+(i-1)R_1+\sum_{j=1}^{i-1}R_j+R_i+(k-i+1)\sigma+(k-i)R_1+\sum_{j=1}^{k-i}R_j+R_{k-i+1}\\&=(k+1)\sigma+(k-1)R_1+\sum_{j=1}^{i}R_j+\sum_{j=1}^{k-i+1}R_j\\&=(k+1)\sigma+kR_1+\sum_{j=2}^{i}R_j+\sum_{j=1}^{k-i+1}R_j\end{aligned}$$

Since $\boldsymbol{R}$ is a non-decreasing sequence, we have

$$\sum_{j=2}^{i}R_j\le\sum_{j=k-i+2}^{k}R_j$$

Then it is true that

$$
\begin{aligned}
& U[i]+U[k-i+1] \\
& \le (k+1)\sigma + kR_1 + \sum_{j=1}^{k-i+1} R_j + \sum_{j=k-i+2}^{k} R_j \\
& = (k+1)\sigma + kR_1 + \sum_{j=1}^{k} R_j \\
& = U[1]+U[k]
\end{aligned}
$$

Thus,

$$L[k+1] = U[1]+U[k] = (k+1)\sigma + kR_1 + \sum_{j=1}^{k} R_j$$

thereby showing that the above equation $L[k+1]$ holds for $t=k+1$. By mathematical induction, Eq. (1) always holds for any integer $t$ ($1 \le t \le T$). Besides, the definition of the upper boundary could easily lead us to Eq. (2). So Theorem 3 is proven. □

Consequently, OPEA is efficient. The average time complexities of the simplified boundary generation, encryption and decryption sub-algorithm are $O(T)$, $O(1)$ and $O(\log T)$, respectively. In particular, **SBoundaryGen** is a one-time function. Private boundaries would not have to be updated after being generated locally. As for some sparse databases, there is no need to calculate partitions for every integer in the plaintext domain. Here, computing on demand is allowed. When a new value appears in the database or a value is updated, the database owner just needs to calculate the boundaries for its corresponding partition, as long as enough ordered random numbers are reserved in the initialization phase.

### 4.4.2 Extended Version of OPEA

According to the converse-negative proposition of Theorem 2 obtained in Section 4.3, the additive numerical relation $a_1+a_2+...+a_k \ge b$ is determined by its ciphertext inequation, i.e., $E(a_1)+E(a_2)+...+E(a_k) \ge E(b)$. But for the negative proposition of Theorem 2, we also study the ciphertext relation $E'(a)+E'(b)>E'(c)$, where $E': X \rightarrow Y'$ is the encryption function in the extension of OPEA. Firstly, an extended boundary generation sub-algorithm (***L'***, ***U'***)←**BoundaryGen'**(***R***, $\sigma$) with the pre-condition $\sigma > 3 \cdot \max_{1 \le i \le T}\{R_i\}$ is designed as the following Algorithm 5. The extended encryption sub-algorithm $E'(m)$←**Enc**($m$, ***L'***, ***U'***) and the extended decryption sub-algorithm $D'(c)$←**Dec**($c$, ***U'***) are performed in the same manner.

**Algorithm 5** (***L'***, ***U'***)←**BoundaryGen'**(***R***, $\sigma$)

1: Set $U'[1]=\sigma$, $L'[1]=U'[1]-R_1$
2: **for** $t$=2 to $T$ **do**
3: Compute $U'[t]=\min_{1 \le i < t}\{L'[i]+L'[t-i]\}$
4: Compute $L'[t]=U'[t]-R_t$, $R_t<U'[t]-U'[t-1]$
5: **end for**
6: **return** ***L'***, ***U'***

Secondly, evidenced by the similar token, Theorem 1 still holds in the extended version and a new conclusion is stated as follows:

**Theorem 4 (Additive order-preservation of extended version).** *For any integers $a+b>c$ ($a$, $b$, $c \in X$), the extended encryption function satisfies that $E'(a)+E'(b)>E'(c)$.*

***Proof.*** This theorem can be proven in exactly the same way as that of Theorem 2, so we have

$$U'[a+b] \le L'[a]+L'[b] \le E'(a)+E'(b)$$

where the first inequation is due to Line 3, Algorithm 5, i.e., $U'[t]=\min_{1 \le i < t}\{L'[i]+L'[t-i]\}$.

And since

$$U'[c]<U'[a+b],\ E'[c] \le U'(c)$$

Then it is true that

$$E'(a)+E'(b)>E'(c)$$

So Theorem 4 follows. The additive numerical relation $a_1+a_2+...+a_k \le b$ can also be determined by its ciphertext inequation $E'(a_1)+E'(a_2)+...+E'(a_k) \le E'(b)$. □

### 4.4.3 Simplified Version of Extended OPEA

$T$+1 non-decreasing random numbers are generated, which satisfy that $R_i \le R_j$ and $\sigma > 3R_T$ for $1 \le i < j \le T$. A simplified version of the extended OPEA (see Algorithm 6) is developed into (***L'***, ***U'***)←**SBoundaryGen'**(***R***, $\sigma$).

**Algorithm 6** (***L'***, ***U'***)←**SBoundaryGen'**(***R***, $\sigma$)

1: Set $S$=0
2: **for** $t$=1 to $T$ **do**
3: Compute $U'[t]=t\sigma-(t-1)R_1-S$
4: Compute $L'[t]=U'[t]-R_t$
5: Compute $S=S+R_t$
6: **end for**
7: **return** ***L'***, ***U'***

It reduces the time complexity of the extended OPEA from $O(T^2)$ to $O(T)$ successfully. From Line 3 to Line 5 in this sub-algorithm, we can also derive a new theorem for the linear boundary function in the extended version.

**Theorem 5 (Linear boundary function of extended version).** *For any integer t* ($1 \le t \le T$), *its extended lower and upper boundaries satisfy that*

$$U'[t] = t\sigma - (t-1)R_1 - \sum_{j=1}^{t-1} R_j \tag{3}$$

$$L'[t] = U'[t] - R_t \tag{4}$$

***Proof.*** This theorem can be proven in exactly the same way as that of Theorem 3. Eq. (4) is easily derived from its definition. The second mathematical induction is used to prove the correctness of Eq. (3).

Basis: In light of Line 1, Algorithm 5, Eq. (3) holds for $t$=1. $U'[1]$ amounts to the equation

$$U'[1]=1\cdot\sigma-(1-1)\cdot R_1=\sigma$$

Inductive step: Let us assume that Eq. (3) holds for any integer $t$ ($1<t\le k$, $k<T$), that is,

$$U'[t] = t\sigma - (t-1)R_1 - \sum_{j=1}^{t-1} R_j, 1 < t \le k$$

Then for $t=k+1$, according to Line 3, Algorithm 5 and the definition of the upper boundary, $U'[k+1]$ can be written as

$$U'[k+1]=\min_{1\le i<k+1}\{L'[i]+L'[k-i+1]\}=\min\{L'[1]+L'[k], \min_{2\le i<k+1}\{L'[i]+L'[k-i+1]\}\}$$

where the left part can be deduced as

$$\begin{aligned} L'[1] + L'[k] &= U'[1] - R_1 + U'[k] - R_k \\ &= \sigma - R_1 + k\sigma - (k-1)R_1 - \sum_{j=1}^{k-1} R_j - R_k \\ &= (k+1)\sigma - kR_1 - \sum_{j=1}^{k} R_j \end{aligned}$$

And for any integer $i$ ($2\le i\le k$), the right part can be deduced as

$$\begin{aligned} &L'[i] + L'[k-i+1] \\ &= U'[i] - R_i + U'[k-i+1] - R_{k-i+1} \\ &= i\sigma - (i-1)R_1 - \sum_{j=1}^{i-1} R_j - R_i + (k-i+1)\sigma - (k-i)R_1 - \sum_{j=1}^{k-i} R_j - R_{k-i+1} \\ &= (k+1)\sigma - (k-1)R_1 - \sum_{j=1}^{i} R_j - \sum_{j=1}^{k-i+1} R_j \\ &= (k+1)\sigma - kR_1 - \sum_{j=2}^{i} R_j - \sum_{j=1}^{k-i+1} R_j \end{aligned}$$

Since $\boldsymbol{R}$ is a non-decreasing sequence, we have

$$\sum_{j=2}^{i} R_j \le \sum_{j=k-i+2}^{k} R_j$$

Then it is true that

$$\begin{aligned} &L'[i] + L'[k-i+1] \\ &\ge (k+1)\sigma - kR_1 - \sum_{j=1}^{k-i+1} R_j - \sum_{j=k-i+2}^{k} R_j \\ &= (k+1)\sigma - kR_1 - \sum_{j=1}^{k} R_j \\ &= L'[1] + L'[k] \end{aligned}$$

Thus,

$$U'[k+1] = L'[1] + L'[k] = (k+1)\sigma - kR_1 - \sum_{j=1}^{k} R_j$$

thereby showing that the equation $U'[k+1]$ holds for $t=k+1$. By mathematical induction, Eq. (3) always holds for any integer $t$ ($1\le t\le T$). So Theorem 5 is proven. □

The next thing brought up for discussion is the effectiveness of partition boundaries. Based on Theorem 5 and the monotonicity of random numbers, two conclusions follow that

$$U'[t] = t\sigma - (t-1)R_1 - \sum_{j=1}^{t-1} R_j \ge t\sigma - (t-1)R_T - (t-1)R_T = (t-1)(\sigma - 2R_T) + \sigma$$

$$L'[t] = U'[t] - R_t = (t-1)(\sigma - 2R_T) + \sigma - R_t \ge (t-1)(\sigma - 2R_T) + (\sigma - R_T)$$

Since $\sigma>3R_T\geq 2R_T\geq R_T\geq 0$, it is true that the ciphertext domain of OPEA belongs to the positive integer field, i.e.,

$$U'[t]>0,\ L'[t]>0,\ 1\leq t\leq T$$

Furthermore, for any integer $t$ ($1<t\leq T$), the interval between two partitions is

$$L'[t]-U'[t-1]=U'[t]-R_t-U'[t-1]=\sigma-R_1-R_{t-1}-R_t\geq\sigma-R_T-R_T-R_T>0$$

Therefore, it can be seen that any two adjacent partitions are non-overlapped, which conforms to the limitation in Line 4, Algorithm 5.

In addition, let us assume that $\varsigma$ ($|\varsigma|\geq 2$) is a set of positive integers, whose arithmetic sum is denoted as SUM($\varsigma$). $E(\varsigma)\leftarrow$**Enc**($\varsigma$, $\boldsymbol{L}$, $\boldsymbol{U}$) and $E'(\varsigma)\leftarrow$**Enc**($\varsigma$, $\boldsymbol{L'}$, $\boldsymbol{U'}$) are two encrypted sets. Thus, for a non-decreasing sequence $\boldsymbol{R}$ and any plaintext (i.e., *value*) with $\sigma\geq(value+|\varsigma|)\cdot R_T-R_1$, there exist two important conclusions on the encrypted sum:

**Theorem 6 (A Lower limit of encrypted sum).** *For any integer value* ($1\leq value<T$), *if* SUM($\varsigma$)>*value*, *then* SUM($E(\varsigma)$)>$L[value]$.

***Proof.*** Firstly, in light of the linear boundary function of the simplified version, the following equality holds

$$\begin{aligned}&\mathrm{SUM}(L[\varsigma])-L[value]\\&=\mathrm{SUM}(\varsigma)\cdot\sigma+\left(\mathrm{SUM}(\varsigma)-|\varsigma|\right)\cdot R_1+\sum_{t\in\varsigma}\sum_{j=1}^{t-1}R_j-value\cdot\sigma-(value-1)\cdot R_1-\sum_{j=1}^{value-1}R_j\\&=(\mathrm{SUM}(\varsigma)-value)\cdot\sigma+\left(\mathrm{SUM}(\varsigma)-|\varsigma|\right)\cdot R_1-\left(value-1\right)\cdot R_1+\sum_{t\in\varsigma}\sum_{j=1}^{t-1}R_j-\sum_{j=1}^{value-1}R_j\end{aligned}$$

Since $\boldsymbol{R}$ is a non-decreasing random sequence in the simplified version, we have that $\max_{1\leq i\leq T}\{R_i\}=R_T$ as well as $\min_{1\leq i\leq T}\{R_i\}=R_1$. So it is further noticed that

$$value\cdot R_T=value\cdot\max_{1\leq i\leq T}\{R_i\}\geq\sum_{j=1}^{value-1}R_j$$

And

$$\begin{aligned}&|\varsigma|\cdot R_T-R_1=|\varsigma|\cdot\max_{1\leq i\leq T}\{R_i\}-R_1\\&\geq\sum_{t\in\varsigma}R_t-R_1=\left(\sum_{t\in\varsigma}\sum_{j=1}^{t}R_j-\sum_{t\in\varsigma}\sum_{j=1}^{t-1}R_j\right)-R_1\\&\geq\mathrm{SUM}(\varsigma)\cdot\min_{1\leq i\leq T}\{R_i\}-\sum_{t\in\varsigma}\sum_{j=1}^{t-1}R_j-R_1\\&>value\cdot\min_{1\leq i\leq T}\{R_i\}-\sum_{t\in\varsigma}\sum_{j=1}^{t-1}R_j-R_1\\&=(value-1)\cdot R_1-\sum_{t\in\varsigma}\sum_{j=1}^{t-1}R_j\end{aligned}$$

Therefore, we have

$$\sigma\geq(value+|\varsigma|)\cdot R_T-R_1>(value-1)\cdot R_1-\sum_{t\in\varsigma}\sum_{j=1}^{t-1}R_j+\sum_{j=1}^{value-1}R_j$$

Then after substituting SUM($\varsigma$)-*value*≥1 and SUM($\varsigma$)≥|$\varsigma$| into the above equality, the SUM($L[\varsigma]$)-$L[value]$>0 follows. Thus, SUM($E(\varsigma)$)≥ SUM($L[\varsigma]$)>$L[value]$. Theorem 6 is completed. □

**Theorem 7 (An upper limit of encrypted sum).** *For any integer value* ($2<value\leq T$), *if* SUM($\varsigma$)<*value*, *then we have* SUM($E'(\varsigma)$)< $U'[value]$.

***Proof.*** This theorem can be proven in exactly the same way as that of Theorem 6, so we have

$$\begin{aligned}&\mathrm{SUM}(U'[\varsigma])-U'[value]\\&=\mathrm{SUM}(\varsigma)\cdot\sigma-\left(\mathrm{SUM}(\varsigma)-|\varsigma|\right)\cdot R_1-\sum_{t\in\varsigma}\sum_{j=1}^{t-1}R_j-value\cdot\sigma+(value-1)\cdot R_1+\sum_{j=1}^{value-1}R_j\\&=(\mathrm{SUM}(\varsigma)-value)\cdot\sigma-\left(\mathrm{SUM}(\varsigma)-|\varsigma|\right)\cdot R_1+\left(value-1\right)\cdot R_1-\sum_{t\in\varsigma}\sum_{j=1}^{t-1}R_j+\sum_{j=1}^{value-1}R_j\end{aligned}$$

It is further noticed that

$$value\cdot R_T\geq\sum_{j=1}^{value-1}R_j$$

And

$$|\varsigma|\cdot R_T-R_1>(value-1)\cdot R_1-\sum_{t\in\varsigma}\sum_{j=1}^{t-1}R_j$$

Therefore, we have

$$\sigma>(value-1)\cdot R_1-\sum_{t\in\varsigma}\sum_{j=1}^{t-1}R_j+\sum_{j=1}^{value-1}R_j$$

Then after substituting SUM(ς)-*value*≥1 and SUM(ς)≥|ς| into the above equality, the SUM(*U*'[ς])-*U*'[*value*]<0 follows. Thus, SUM(*E*'(ς)) ≤SUM(*U*'[ς])<*U*'[*value*]. Theorem 7 is completed. □

The analyses of functionality limitations in OPEA indicate that only the comparison between a summation of a set of ciphertexts and a fixed threshold is available (for more illustrations, please see Section 4.6). So there may be some possible extensions in functionalities and one would also like to present a novel additive order preserving encryption scheme. Related cryptographic techniques need more investigation in the future.

## 4.5 Security Analysis

Formal analyses in this subsection demonstrate that our OPEA can resist ciphertext-only attacks, statistical attacks and weak chosen-plaintext attacks in the process of executing data manipulations.

### *4.5.1 Ciphertext-Only Attack*

Due to the undiscovered partition precision, the initial location $\sigma$ as well as the partition length $\boldsymbol{R}$, the adversary *A* fails to infer the plaintext distributions based on the distance between ciphertexts. If *A* is allowed to select a finite number of ciphertext values in the game, he should first determine the lower and upper boundaries of the encrypted partition $[L_m, U_m]$ in order to guess successfully the plaintext *m* and its secret key $R_m$. And for this purpose, *A* needs to get at least four encrypted values (i.e., $V_1$, $V_2$, $V_3$ and $V_4$), where $V_1$ is the ciphertext of *m*-1, $V_2$ and $V_3$ are the ciphertexts of *m*, and $V_4$ is the ciphertext of *m*+1. Meanwhile, $V_2=V_1+\mu_1$ and $V_4=V_3+\mu_2$ should be set up, such that $\mu_1$ and $\mu_2$ are the interval length between two partitions. From the inequation $L[t]-U[t-1]\geq U[1]=\sigma+R_1$, $1<t\leq T$, one could easily argue that $\mu_1$ and $\mu_2$ are influenced by the initial random number $\sigma+R_1$. Besides, a ciphertext $E(m)$ is randomly selected from the partition $[L_m, U_m]$ with length $R_m=U_m-L_m$, then the probability of successful attack is at most

$$P[\text{success}^{A,\boldsymbol{R},\sigma}] \leq \left(\frac{1}{\sigma+R_1}\right)^2 \cdot \frac{1}{R_{m-1}} \cdot \frac{1}{R_{m+1}}$$

The justification is that, in the one-to-many mapping, every encryption is mapped to different ciphertexts, and every decryption does not provide any assistance to analyze the private boundaries. It can be seen that the bigger $\sigma+R_1$ or $R_m$ is, the wider interval or alternative ciphertext partition will be. So by choosing the random parameters appropriately, the $P[\text{success}^{A,\boldsymbol{R},\sigma}]$ tends to be negligible, and the security is enhanced significantly with the help of randomness in this probabilistic encryption. Hence, we can conclude that OPEA is secure enough under a ciphertext-only attack.

### *4.5.2 Statistical Attack*

The statistical method of calculating ciphertext frequency is also one of the possible ciphertext-only attacks. Under this attack, if two plaintext values ($m_1$ and $m_2$) have a totally different distribution like $P[x=m_1]<P[x=m_2]$, then the adversary *A* may assert that their ciphertext values should satisfy that $P[x=E(m_1)]<P[x=E(m_2)]$ with a high probability. As a consequence, he successfully breaks the cryptosystem. In fact, most of the existing OPE schemes are vulnerable to statistical attacks because of their deterministic one-to-one mapping.

Probabilistic encryption is used in OPEA to prevent the statistical attack. The plaintext distribution is altered by the one-to-many mapping, so the probability of ciphertext distributions turns to be

$$P[y=E(m)]=\frac{P[x=m]}{R_m}$$

By this way, OPEA succeeds in preventing the statistical attack. A simulative evaluation was performed on the input and output distribution of OPEA. Figure 4 shows the distributions of ciphertext domain under different distributions of plaintext domain (e.g., normal distribution or uniform distribution).

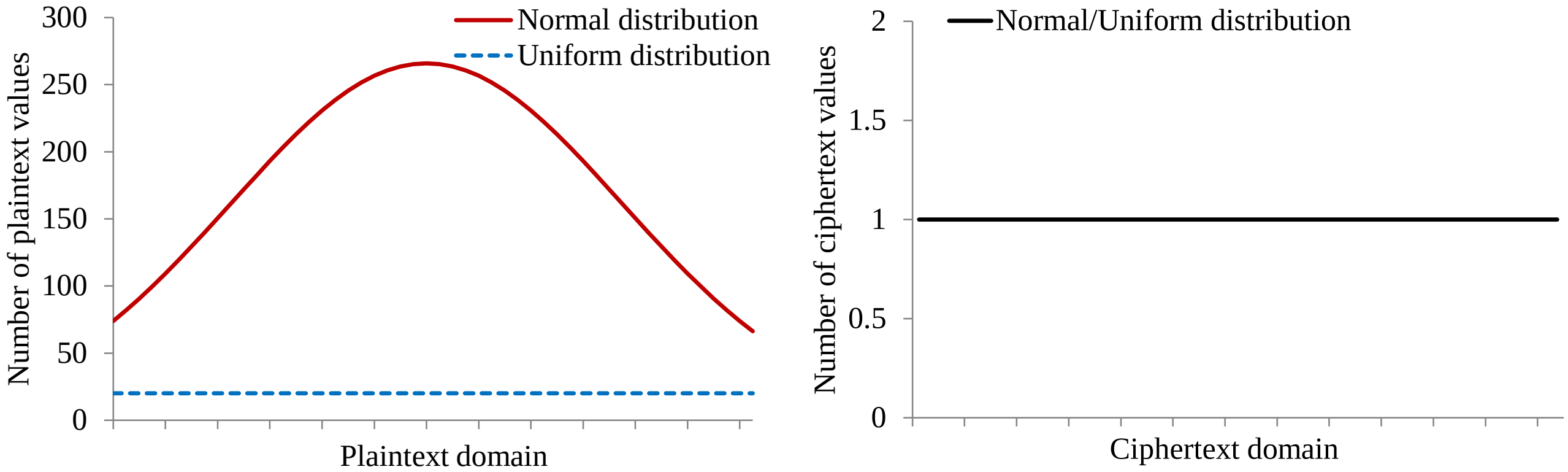

**(a) Plaintext domain distribution** **(b) Ciphertext domain distribution**

**Figure 4: Input and output distribution of OPEA.**

As shown in Figure 4, it is obvious that the ciphertext domain encrypted by OPEA is nearly uniform (Figure 4b) for bigger $R_m$, regardless of the plaintext domain distribution (Figure 4a), which suggests that the adversary *A* cannot acquire any additional statistical information from ciphertexts.

### *4.5.3 Chosen-Plaintext Attack*

Theoretically, if it is assumed that a malicious attacker has some prior knowledge about finite plaintext values, all of the existing OPE schemes based on one-to-one mapping are not secure for the known-plaintext attack. For example, if an attacker obtains two pairs of given plaintext-ciphertext, that is, ($m_1$, $c_1$) and ($m_2$, $c_2$), then as for another ciphertext $c_3$ with $c_1<c_3<c_2$, the attacker will easily guess that its corresponding plaintext value $m_3$ belongs to the interval [$m_1$, $m_2$]. As you could see above, knowing that specific datasets have been encrypted with OPE provides useful information to get the approximate plaintext value. Clearly, such OPE schemes are also vulnerable to a chosen-plaintext attack in the standard model.

So, one approach to guarantee security for OPEA is to try to reveal no additional information about the plaintext values except their numerical order. The most ideal security definition is INDistinguishability under Ordered Chosen-Plaintext Attack (IND-OCPA), presented by Boldyreva et al. [27]. Surprisingly, almost no OPE scheme achieves it. Popa et al. [12] have proven that any stateful IND-OCPA-secure OPE must have ciphertext size at least exponential in the plaintext size, and the ciphertext mutability is necessary for IND-OCPA security. Regarding our OPEA, although it is a stateful algorithm with mutable ciphertexts, its additivity leaks the multiple relations or the distance relations among plaintexts, causing vulnerabilities in IND-OCPA.

For simultaneous order and additivity preservation, we weaken the IND-OCPA [27] security notion to the INDistinguishability under an Additive Ordered Chosen-Plaintext Attack (IND-AOCPA), and show that the honest-but-curious cloud servers cannot obtain any additional sensitive information except that inferred by the ciphertexts' order and additivity. In a practical sense, this information leakage is inevitable to guarantee the normal data manipulation of encrypted databases.

As for the security game for IND-AOCPA, it requires the chosen plaintexts in each round to satisfy the same order and multiple relations. In what follows, let us suppose that the CSP does not have any prior knowledge about the plaintext domain, and there is no collusion in any way between the CSP and database users. Queries on sequential plaintext values are generally not allowed, and the pseudo-random number generator used to assign secret keys is assumed to be secure.

Similar to [12], the following game is modeled between the database owner and a Probabilistic Polynomial Time (PPT) adversary *A* from the CSP, which is called **IND-AOCPA Game**. It has one more constraint to the adversary *A* compared to previous IND-OCPA Game, i.e., the plaintexts in queries are bounded by a proportional relation. Our **IND-AOCPA Game** is consisting of three stages.

1) The database owner generates the private boundaries ($\boldsymbol{L}$, $\boldsymbol{U}$) by secret keys ($\boldsymbol{R}$, $\sigma$), and selects a random bit *b*.

2) The database owner and the adversary *A* repeat an adaptive encryption query through a polynomial number of rounds. Then at the round *i*:

a. The adversary *A* selects two values $v^0(i)$ and $v^1(i)$ from the plaintext domain, and sends them to the database owner.

b. The database owner encrypts the value $v^b(i)$ by **Enc**($v^b(i)$, $\boldsymbol{L}$, $\boldsymbol{U}$), and sends its ciphertext $E(v^b(i))$ to the adversary *A*.

3) The adversary *A* outputs his guess *b*'.

The adversary *A* wins in this security game when the following three conditions hold simultaneously:

1) The adversary *A* outputs the right guess, namely, *b*=*b*'.

2) The input sequences $\{v^0(i)\}$ and $\{v^1(i)\}$ have the same order relations, namely, for any *i*, *j*, $v^0(i)<v^0(j)$ if and only if $v^1(i)<v^1(j)$.

3) The input sequences $\{v^0(i)\}$ and $\{v^1(i)\}$ have the same multiple relations, namely, for any *i*, *j*, $x_i \subseteq R^+$-{1}, $v^0(j)=x_i \cdot v^0(i)$ if and only if $v^1(j)=x_i \cdot v^1(i)$.

Then we have a new security definition.

**Definition 1.** An OPE is IND-AOCPA-secure if any polynomial time adversary has only a negligible advantage in the **IND-AOCPA Game** shown above, where the adversary wins with a probability $P_{cpa}[\text{win}^{A,\boldsymbol{R},\sigma}]$ and his advantage is defined as

$$Adv_{cpa}(A) := P_{cpa}[\text{win}^{A,\boldsymbol{R},\sigma}] - \frac{1}{2}$$

**Theorem 8.** *OPEA is IND-AOCPA-secure*.

***Proof.*** The main step is to prove the information that *A* obtains in the case *b*=0 is the same as that in the case *b*=1. Assume by inductive method that before the second game stage and after the *i*th round in the second stage, the same information is learned by *A*. Then after the (*i*+1)th round encryption, since the sequences $\{v^0(i)\}$ and $\{v^1(i)\}$ have the same multiple relations, $v^0(i+1)$ and $v^1(i+1)$ satisfy that

$$v^0(i+1) = x_i \cdot v^0(i) = x_i \cdot x_{i-1} \cdot v^0(i-1) = \cdots = x_i \cdot x_{i-1} \cdots x_1 \cdot v^0(1) = \prod_{t=1}^{i} x_t \cdot v^0(1)$$

$$v^1(i+1) = x_i \cdot v^1(i) = x_i \cdot x_{i-1} \cdot v^1(i-1) = \cdots = x_i \cdot x_{i-1} \cdots x_1 \cdot v^1(1) = \prod_{t=1}^{i} x_t \cdot v^1(1)$$

Notice that

$$\sum_{t=1}^{\omega_1} v^0(t) = v^0(1) + v^0(2) + \cdots + v^0(\omega_1) = \sum_{t=1}^{\omega_1} \prod_{s=1}^{t-1} x_s v^0(1),\ 0 < \omega_1 \le i+1$$

$$\sum_{t=1}^{\omega_2} v^1(t) = v^1(1) + v^1(2) + \cdots + v^1(\omega_2) = \sum_{t=1}^{\omega_2} \prod_{s=1}^{t-1} x_s v^1(1),\ 0 < \omega_2 \le i+1$$

Then it is true that

$$\sum_{t=1}^{\omega} v^0(t) < v^0(i+1) \ \boldsymbol{iff} \ \sum_{t=1}^{\omega} v^1(t) < v^1(i+1),\ 0 < \omega \le i+1$$

So the adversary $A$ obtains nothing from ciphertexts other than the same duplicates, multiple relations and order relations in both cases. In the point of information theory, there is no way for $A$ to distinguish between $\{v^0(i)\}$ and $\{v^1(i)\}$. For all sufficiently large secret keys ($\boldsymbol{R}$, $\sigma$), a conclusion follows that

$$P_{cpa}[\text{win}^{A,\,\boldsymbol{R},\,\sigma}] \le \frac{1}{2} + \text{negl}(\boldsymbol{R}, \sigma)$$

where negl($\boldsymbol{R}$, $\sigma$) is a negligible probability. The proof of Theorem 8 is completed. □

### 4.5.4 Chosen-Ciphertext Attack

Similar to Section 4.5.3, we also weaken the INDistinguishability under Chosen-Ciphertext Attack (IND-CCA for short) appropriately by making assumptions about attacks to a new notion called INDistinguishability under Ordered Chosen-Ciphertext Attack (IND-OCCA for short). The IND-OCCA has one more constraint to the adversary $A$ compared to previous IND-CCA, i.e., it requires the messages chosen in the challenge stage to have the same order relation as the plaintexts decrypted in the find stage.

The IND-OCCA security game is modeled between the database owner and a stateful PPT adversary $A$, which is called **IND-OCCA Game** and is consisting of four stages.

1) Setup stage: The database owner generates the private boundaries ($\boldsymbol{L}$, $\boldsymbol{U}$) by secret keys ($\boldsymbol{R}$, $\sigma$) and selects a random bit $b$.

2) Find stage: The database owner and the adversary $A$ perform a polynomial number of decryption queries as follows.

a. The adversary $A$ selects an encrypted value $c=E(m)$ and sends it to the database owner.

b. The database owner decrypts the encrypted value $E(m)$ by **Dec**($c$, $\boldsymbol{L}$, $\boldsymbol{U}$) and sends its plaintext $m$ to the adversary $A$.

3) Challenge stage: The database owner and the adversary $A$ perform an adaptive encryption query in polynomial time.

a. The adversary $A$ selects a pair of plaintexts ($m_0$, $m_1$) from the plaintext domain and sends it to the database owner.

b. The database owner encrypts the plaintext value $m_b$ by **Enc**($m_b$, $\boldsymbol{L}$, $\boldsymbol{U}$) and sends its ciphertext $E(m_b)$ to the adversary $A$.

4) Output stage: The adversary $A$ outputs his guess $b$'.

The adversary $A$ wins in the above security game with the probability $P_{cca}[\text{win}^{A,\,\boldsymbol{R},\,\sigma}]$ when the following two conditions hold simultaneously:

1) The adversary $A$ outputs the right guess, namely, $b=b'$.

2) The challenge pair ($m_0$, $m_1$) has the same order relation as the plaintext $m$, namely, $m_0<m$ if and only if $m_1<m$.

**Definition 2.** An OPE algorithm is IND-OCCA-secure if any polynomial time adversary has only a negligible advantage in **IND-OCCA Game** shown above, where the adversary's advantage is defined as

$$Adv_{cca}(A) := P_{cca}[\text{win}^{A,\,\boldsymbol{R},\,\sigma}] - \frac{1}{2}$$

**Theorem 9.** *OPEA is not IND-OCCA-secure.*

***Proof.*** Consider the following adversary $A$ against our OPEA algorithm:

Without loss of generality, let us assume that $m_0<m_1$. Then there exists a real number $n$ ($m_0/m<n<m_1/m$), s.t., $nE(m)<E(m_1)$. Therefore, the adversary $A$ chooses $n$ in the challenge stage and judges the output value on the strategy that

***if*** $nE(m)\ge E(m_b)$ ***then*** Output 0 ***else*** Output 1

And we claim that $P[b'=0|b=0]>1/2$ and $P[b'=1|b=1]=1$. For all sufficiently large secret keys ($\boldsymbol{R}$, $\sigma$), the probability that $A$ wins the **IND-OCCA Game** is that

$$P_{cca}[\text{win}^{A,\,\boldsymbol{R},\,\sigma}]=|P[b'=1|b=1]-P[b'=1|b=0]|=1-(1-P[b'=0|b=0])=P[b'=0|b=0]>1/2$$

With further analyses, OPEA is clearly not IND-OCCA-secure. It is not secure under the ordered chosen-ciphertext attack as long as a plaintext-ciphertext pair is revealed by $A$. It is no longer important to consider the multiple relations now. The proof of Theorem 9 is completed. □

For T-DB system to simultaneously perform range queries and aggregation queries over ciphertexts, the IND-AOCPA in the OPEA algorithm is the best security that has been proven formally so far.

Besides, as [30] has noted, there are a series of ciphertext-only inference attacks on the relational databases encrypted using OPE. According to the research conclusions [30] from Naveed et al., given the encrypted database along with some auxiliary information, an attacker can guarantee the success of recovering parts of underlying plaintext values. Thus it can be seen that the order-preservation makes OPE algorithms suitable for range queries on encrypted data, while it also inherently limits the achievable security. To this end, a recent research team [31, 32] from the Stanford University constructed a highly-efficient, non-interactive and stateless Order-Revealing Encryption (ORE) scheme with the best-possible semantic security. It takes advantage of the special structure of ciphertexts to perform ciphertext search while protect against inference attacks. However, before we use the ORE in place of OPE to support range queries in the encrypted database, the cloud RDBMS is required to deploy a custom comparator. In addition, since the ORE does not include any decryption function, its encryption function and comparison function have to be invoked many times to decrypt the query results. The final plaintexts are recovered by a binary search, which causes a significant increase in the computational overhead of the database owner. In conclusion, ORE techniques are far from being practically viable in the encrypted database, and their supported query functionalities are also limited so far. Moreover, the study of a secure ORE with additivity is still an open and notable question.

## 4.6 Algorithm Limitation

As the cipher tool in T-DB, OPEA settles the problem about the utilization of data manipulations in outsourced encrypted databases. Nevertheless, there are still some deficiencies remaining. For example, combined with the additive order-preservations in Theorem 2 and Theorem 4, a corollary is made as follows:

**Corollary 1.** *There is no such probabilistic OPE scheme that can be used to judge the order relation between the sum expression* $a_1+a_2+...+a_k$ *and* $b_1+b_2+...+b_{k'}$.

***Proof.*** The proof is by contradiction. Firstly, we assume that there exists a probabilistic OPE algorithm to compare $E(a_1)+E(a_2)+...+E(a_k)$ and $E(b_1)+E(b_2)+...+E(b_{k'})$, that is to say, the lower and upper boundary of the partition $t$ satisfy that $L[t]=\max_{1\le i<t}\{U[i]+U[t\text{-}i]\}$ and $U[t]=\min_{1\le i<t}\{L[i]+L[t\text{-}i]\}$ simultaneously.

Then, we could find that

$$U[a]+U[b]\le L[a+b]\le U[a+b]\le L(a)+L(b)$$

It is easy to see that since no overlap exists between different partitions, the above expression $U[a]+U[b]\le L[a]+L[b]$ holds if and only if

$$U[a]+U[b]=L(a)+L(b)$$

which is contrary to our assumption that the OPE is a probabilistic algorithm. □

Hence, the above assumption is actually invalid. There is no feasible method to construct a probabilistic encryption algorithm by dividing ciphertext partitions with intervals, otherwise the order-preservation and the additive homomorphism should be satisfied simultaneously, which turns OPEA into a completely deterministic encryption.

# 5. TRANSLATION MODULE: SQL-TRANSLATOR

The translation module works to hide the outsourced databases' changes that result from encryption from the CSP and database users so that they can behave as if the database were not encrypted. The database users can query the database server as usual and receive the appropriate answers. The CSP can process almost all kinds of SQL queries (from the database owner) as in the standard DBaaS pattern, and may not notice that it is executing manipulations on an encrypted database. We work on Microsoft SQL Server but stress that our work can be easily extended to other relational databases.

The general method adopted by T-DB system to execute data manipulations over an outsourced encrypted database is discussed in this section. To begin with, two user-defined functions are designed to solve the problem of equality comparison. Secondly, an SQL interpreter (called SQL-Translator) is deployed on the database owner to translate a plaintext SQL statement into ciphertext SQL statements that can be executed directly on the cloud server. Various translation rules about common data manipulations and relational operations [33] in the process of translating statements are introduced in detail. We finally summarize several classes of functionality limitations and formal syntax rules in SQL-Translator.

## 5.1 Equality Comparison Function

Two cloud UDFs are established as follows:

1) Equality comparison for two values (Algorithm 7).

Since OPEA is a kind of probabilistic encryption algorithm, the same plaintext may be encrypted to different ciphertexts frequently. In order to determine whether two plaintexts ($value_1$ and $value_2$) are equal by their ciphertexts (i.e., $E(value_1)$ and $E(value_2)$), the database owner should pick out a random integer $x\in[\max\{R_1, R_2\}, \sigma+R_1)$ based on secret keys and send it to the cloud RDBMS, where $R_1$ (resp. $R_2$) is the length of the partition that $value_1$ (resp. $value_2$) belongs to. As we can see from Line 3, Algorithm 1, the expression $L[t]\ge U[1]+U[t\text{-}1]$ holds for any $t$ ($t$=2, 3, …, $T$), so the interval between two sequential partitions satisfies that

$$L[t]\text{-}U[t\text{-}1]\ge U[1]=\sigma+R_1>x\ge\max_{1\le i\le T}\{R_i\}\ge R_j,\ 1\le j\le T$$

That is to say, the $x$ is between the partition length $R_j$ ($1\le j\le T$) and the interval length $L[t]$-$U[t$-1] ($t$=2, 3, …, $T$). The difference between ciphertexts of two different plaintexts is bigger than $x$, while the difference between ciphertexts of a single plaintext is no bigger than $x$. Hence, a cloud UDF (i.e., **EqualityCom**) for two-value equality comparison can be made by the pre-selected $x$.

**Algorithm 7 EqualityCom**($x$, $E(value_1)$, $E(value_2)$)

**if** $|E(value_1)\text{-}E(value_2)|\le x$ **then**
  **return** 0 //$value_1=value_2$
**else if** $E(value_1)>E(value_2)$ **then**
  **return** 1 //$value_1>value_2$
**else**
  **return** -1 //$value_1<value_2$
**end if**

This algorithm is running on the random number $x$ and two encrypted values. If output is 0, then two plaintext values are equal; If output is 1, then the first plaintext value is bigger than the second one; If output is -1, then the first plaintext value is less than the second one.

In general, the selection of $x$ is based on some partition boundaries of the ciphertexts that are to be queried by the database users, thereby avoiding the possibility that an attacker determines the equality of other plaintexts. Every time the database user proposes a new query request, the database owner will reselect $x$ according to the encryption granularity and scope of data items. This value is only available for all the possible plaintext items requested by the current database user within his authorized scope, which does not reveal more information about the outsourced data on the cloud. The CSP would invoke the **EqualityCom** function to accomplish each subsequent equality comparison in pairs, without involving the database owner.

2) Equality comparison for arithmetic sum (Algorithm 8).

This part focuses on the equality comparison between an arithmetic sum (i.e., SUM($\varsigma$), $|\varsigma|\ge2$) and a plaintext (i.e., *value*).

By Theorems 6 and 7, if SUM($E(\varsigma)$)$\le L[value]$ and SUM($E'(\varsigma)$)$\ge U'[value]$ hold simultaneously, then we can infer that

$$\text{SUM}(\varsigma)\le value,\ \text{SUM}(\varsigma)\ge value$$

that is, SUM($\varsigma$)=*value*. Hence, a cloud UDF for aggregation comparison (i.e., **SumEqualityCom**) can be made.

**Algorithm 8 SumEqualityCom**(SUM($E(\varsigma)$), SUM($E'(\varsigma)$), $L[value]$, $U'[value]$)

**if** SUM($E(\varsigma)$)$\leq L[value]$&SUM($E'(\varsigma)$)$\geq U'[value]$ **then**
  **return** 0   //SUM($\varsigma$)=*value*
**else if** SUM($E'(\varsigma)$)$>U'[value]$ **then**
  **return** 1   //SUM($\varsigma$)>*value*
**else if** SUM($E(\varsigma)$)$<L[value]$ **then**
  **return** -1   //SUM($\varsigma$)<*value*
**end if**

This algorithm runs securely on the CSP. If output is 0, then the arithmetic sum is equal to the plaintext value; If output is 1, then the arithmetic sum is bigger than the plaintext value; If output is -1, then the arithmetic sum is less than the plaintext value.

## 5.2 Design Principle

The database owner is in charge of translating plaintext SQL statements, so the process of data manipulations is transparent to database users and the CSP. The cloud server can execute exact data manipulations over outsourced encrypted databases without being able to find out the original request. Different from previous studies [22, 23], the key point of SQL translation in T-DB lies in how to run all kinds of data manipulations on the results of a single encryption.

A syntax interpreter based on standard Transact-SQL Data Manipulation Language (DML) [34] is implemented to translate a plaintext SQL statement into a set of ciphertext SQL statements with negligible computational costs. It targets the outsourced database that is stored in the form of OPEA ciphertext. Some specific rules of storage format can be obtained from Section 7.3. SQL-Translator enables various types of data *insert*, *delete*, *update* and *query* operations through one-time encryption. It mainly includes equality query (=; Equi-Join; IS; IN; GROUP BY), range query (>; <; ORDER BY; BETWEEN AND), aggregation query (MIN; MAX; COUNT; SUM; AVG), fuzzy query (LIKE), etc., all of which meet the basic requirements of database manipulations.

The major procedure of SQL-Translator is described as Figure 5, where a minimum translation unit is defined as a sub-statement that contains a perfectly formed clause structure or predicate expression, i.e., a separate operator and data operational entities.

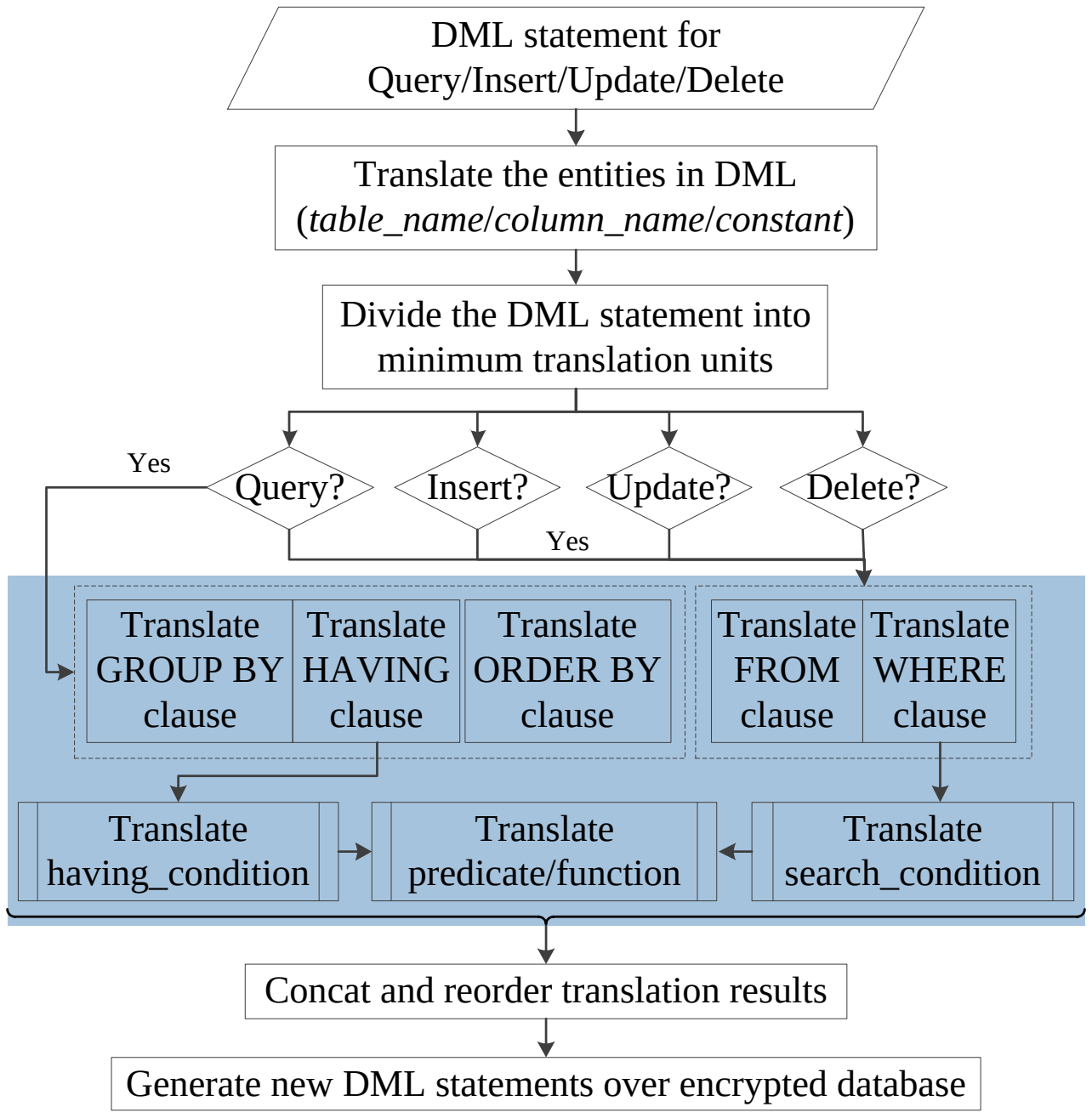

**Figure 5: Generalized process of SQL-Translator.** Blue shading indicates core translations and the sub-modules it calls.

Specific translation steps are as follows:

1) Extract and then translate database entities in the plaintext DML statement, including table names, data fields, constants, etc. The translation rule is that

a. The table name or the data field is translated with a symmetric encryption algorithm (e.g., AES, Blowfish, etc.) or a salted collision-resistant hash function for anonymization.

b. The constant in the statement is translated by the OPEA algorithm, e.g., the constant expression in a range query, the pattern string in a fuzzy query (except the wildcard characters), and the data items in *insert*, *delete* and *update* operations.

2) Divide the statement by SQL keywords and syntactic structure into minimum translation units.

3) Invoke the corresponding translation module in turn to translate the minimum translation unit, see Section 5.3 and Section 5.4 for more details about translation rules.

4) Concat and reorder the translation results, while preserving DML semantics.

5) Return the ciphertext DML statements that can be executed on the outsourced encrypted database directly.

The set of ciphertext SQL statements translated by SQL-Translator strictly follows Transact-SQL standard, so it runs directly on the universal RDBMS. If there is no need for the ciphertext manipulation to declare variables dynamically, then the translated results might be expressed as a store procedure to improve the query efficiency of cloud databases.

To facilitate the understanding of SQL-Translator, we take the following plaintext query statement as an example:

$$\text{SELECT } Att_1 \text{ FROM } tbl \text{ WHERE } Att_1{=}Att_2 \text{ ORDER BY } Att_1 \text{ GROUP BY } Att_1 \text{ HAVING SUM}(Att_2){>}num$$

where *tbl* is a table name, $Att_1$ and $Att_2$ are the column names, and *num* is a constant. The above plaintext statement is translated by SQL-Translator into the following ciphertext query statement that can be executed directly over the encrypted database:

$$\text{SELECT } Att_1^C \text{ FROM } tbl^C \text{ WHERE } \mathbf{EqualityCom}(x, Att_1^C, Att_2^C){=}0 \text{ ORDER BY } Att_1^C$$
$$\text{GROUP BY } Att_1^C \text{ HAVING } \mathbf{SumEqualityCom}(\text{SUM}(E(Att_2)), \text{SUM}(E'(Att_2)), L[num], U'[num]){>}0$$

where $*^C$ denotes the corresponding database object or data manipulation in the ciphertext domain. $L[num]$ and $U'[num]$ represent relevant judgment parameters about *num*. **EqualityCom** and **SumEqualityCom** are two UDFs which are built by the cloud in advance (Algorithms 7 to 8 and Protocol 1). They are invoked by the CSP to make equality comparison or aggregation comparison respectively (see Section 5.1 for more details about UDFs). The $x$ value is a pre-selected random number used to judge the equality in our probabilistic encryption algorithm.

In what follows, we illustrate various translation rules of minimum translation units. More formal syntax conventions and syntax rules are available at [35] and Section 5.6 for lack of space.

## 5.3 Translation Rule of Data Manipulation

The detailed translation rules of minimum translation units (extended functionalities may always be simplified or decomposed to these units) are published in this subsection. A data manipulation statement $s$ over the plaintext database $DB$ is translated into ciphertext statements $s^C$ over the encrypted database $DB^C$. For brevity, in the rest of this paper, $Att$, $Att_1$ and $Att_2$ denote field names in a plaintext table, and $val$, $val_1$ and $val_2$ denote three constants.

### *5.3.1 Query*

A query manipulation is to retrieve the data tuples that satisfy certain query conditions. SELECT is the most complex statement in the plaintext database, whose basic grammatical structure is:

$$\text{SELECT } \langle Att_1, Att_2, \ldots\rangle \text{ FROM } \langle table_name\rangle$$

In the encrypted database, the above statement is translated into

$$\text{SELECT } \langle Att_1^C, Att_2^C, \ldots\rangle \text{ FROM } \langle table_name^C\rangle$$

where the equality of $Att_1^C$, $Att_2^C$ and $table_name^C$ can be judged directly, because they are all encrypted by a deterministic encryption. The columns requested by the database user will be returned to the database owner from the CSP, and no additional operation is needed after decryption. Besides the FROM clause where various forms of JOIN operators (i.e., INNER JOIN, RIGHT JOIN, LEFT JOIN and FULL JOIN) are supported, there are some other clauses in SELECT statement, as described next.

1) WHERE clause.

Regular WHERE includes a comparison predicate, which specifies one or more query conditions to restrict the tuples returned by the query. Common operators and conditions are indicated as follows, where $x$ is a random number in $[\max_{1\le i\le T}\{R_i\}, \sigma{+}R_1)$.

a. Comparison operator ($\theta$).

Operator $\theta$ is usually divided into three kinds of basic translation rules, as illustrated below:

A query condition like '$Att_1{=}Att_2$' is translated into

$$\mathbf{EqualityCom}(x, Att_1^C, Att_2^C){=}0$$

A query condition like '$Att_1{>}Att_2$' is translated into

$$\mathbf{EqualityCom}(x, Att_1^C, Att_2^C){>}0$$

A query condition like '$Att_1{<}Att_2$' is translated into

$$\mathbf{EqualityCom}(x, Att_1^C, Att_2^C){<}0$$

Similar translations may be easily adapted for other comparison operators, such as <>, !=, !>, <=, !<, >=, etc.

b. BETWEEN operator.

Operator BETWEEN specifies an inclusive scope. A query condition like '$Att$ BETWEEN $Att_1$ AND $Att_2$' is translated into

$$\mathbf{EqualityCom}(x, Att^C, Att_1^C){>=}0 \wedge \mathbf{EqualityCom}(x, Att^C, Att_2^C){<=}0$$

c. IN operator.

Operator IN is utilized on an enumerated set to judge whether an attribute value is equal to one of the multiple possible values. A query condition like '$Att$ IN ($Att_1$, $Att_2$, …)' is translated into

$$\mathbf{EqualityCom}(x, Att^C, Att_1^C){=}0 \vee \mathbf{EqualityCom}(x, Att^C, Att_2^C){=}0 \vee \ldots$$

d. IS operator.

Operator IS can find out the tuples with NULL or empty attribute value. NULL signifies that the value is missing or unknown, which is allowed to be consistent in both the plaintext database and the encrypted database. A query condition like '$Att$ IS NULL' is translated into

$$\mathbf{EqualityCom}(x, Att^C, \text{NULL}^C){=}0$$

Similar translations may be easily adapted for another operator like NOT BETWEEN, NOT IN or IS NOT.

e. LIKE operator.

Operator LIKE works to carry out a wildcard-based fuzzy query. It requires that the query column to be stored as ASCII characters (e.g., char, varchar, text, etc.) or some related convertible data types. This column should follow specific encryption rules, as described in Section 7.3. SQL-Translator supports all four patterns (i.e., %, _, [] and [^]) as well as the escaping of wildcard or special literal characters. Its basic grammatical structure is shown as below:

*Att* [NOT] LIKE *pat* [ESCAPE *esc*]

where *pat* is a search pattern string and *esc* is an escape character.

Before invoking LIKE translation module, a table-valued UDF (e.g., **Split**(@str, @delimiter), please see source code [75] for its SQL definition) is built on the cloud RDBMS. An encrypted string is divided by the delimiter and is returned in the form of a Table type. Besides, a temporary column *Att_Match* is added to the field *Att*, in order to save the final matching results of the encrypted string. The internal structure of *pat* is analyzed, and regular characters are escaped and encrypted simultaneously.

Taking the encrypted search pattern string '*αα*%*ββ*%*γγ*' as an example, translating LIKE operator involves several steps: Firstly, we split *pat* by the wildcard character % into three parts (i.e., *αα*, *ββ* and *γγ*). Then a double-layer cursor is used in matching. The first cursor is in charge of traversing the encrypted column in the LIKE clause from top to bottom. It tends to determine whether the length of data items in the current line conforms to the minimal matching length of *pat*. The second cursor sequentially matches the beginning, the middle as well as the ending part of the current encrypted string from left to right. So in other words, it checks for the following three matching conditions in sequence: (i) The current encrypted string begins with *αα*; (ii) and the current encrypted string ends with *γγ*. Here, automatic filtration and matching are also included for trailing blanks. (iii) In addition to the beginning and the ending, the current encrypted string still contains the substring *ββ*. Moreover, AND or OR is utilized to connect the matching conditions in the above. Afterwards, we update the temporary column *Att_Match* and the SQL query condition according to the matching results of the current encrypted string.

The whole process of the above matching is done in ciphertext domain, but there are also some limitations in the translated set of the ciphertext SQL statements. For instance, encrypting the string by splitting into characters leaks the string length; using double-layer cursor and SQL loop statement reduces the query efficiency of wildcard characters.

f. Subquery.

The subquery in SELECT statement mainly falls into the one with EXISTS and the one with a comparison operator. Standard Transact-SQL allows of 32 nesting levels or less.

SQL-Translator translates the nested query level by level. We take the double-level nested query '$s_1$ WHERE EXISTS ($s_2$)' as an example, where $s_1$ and $s_2$ are two basic SELECT statements. The inner query $s_2$ is first translated and its query results are stored into a corresponding temporary intermediate table (e.g., *#INTER_ TABLE2*), then the outer query will be translated into

$s_1$ WHERE EXISTS (SELECT * FROM *#INTER_TABLE2*)

Similar translation may be easily adapted for the subquery with comparison operator. The main difference is that the query results of the inner query will be stored into the corresponding temporary intermediate variables. Other statements with more nesting levels can be translated iteratively.

g. Composite condition.

Multiple query conditions constitute a composite condition by Boolean operators (e.g., AND, OR, etc.) among predicate expressions. As for two query conditions (i.e., $con_1$ and $con_2$), a composite condition like '$con_1$ AND $con_2$' is translated into

$$con_1^C \wedge con_2^C$$

A composite query condition like '$con_1$ OR $con_2$' is translated into

$$con_1^C \vee con_2^C$$

Thus it can be seen that a composite condition is translated directly by concatenating each condition after translating individually. For the above operators (i.e., a to e), their translation rules are nearly identical when *Att*, $Att_1$ or $Att_2$ is a constant.

2) ORDER BY clause.

The ascending or the descending ORDER BY clause sorts the tuples in query results by specified fields. An ORDER BY clause like 'ORDER BY $Att_1$, $Att_2$, … {ASC | DESC}' is translated into

ORDER BY $Att_1^C$, $Att_2^C$, … {ASC | DESC}

After sorting through the first round, **EqualityCom**(·) is used to judge whether there are two or more duplicative tuples with equal encrypted attribute value $Att_1^C$. If so, they are sorted by the subsequent encrypted data field $Att_2^C$.

3) GROUP BY clause.

Grouping statistics are implemented by the GROUP BY clause. It projects the tuples with the same attribute value into a group. For a GROUP BY clause like 'GROUP BY *Att*', we first execute a self-join query on the data column *Att*. Then the query results that satisfy WHERE conditions are stored in a temporary table *#TEMPORARY_TABLE*, resulting that the encrypted attribute values in the same group are completely equal. After that, the grouping is translated into an operation on the temporary table, that is,

GROUP BY $Att^C$

4) HAVING clause.

The HAVING clause is used to further filter the groups resulting from the GROUP BY clause. In addition to the query conditions mentioned above, this clause supports some aggregate functions. In general, these functions can replace the table fields (e.g., *Att*, $Att_1$ or $Att_2$) to participate in the query. The translation rules of different aggregate functions are defined as follows:

a. MIN function.

The aggregate function MIN(·) returns the minimum attribute value that matches the query condition or that consists in a group. An aggregation operation like 'MIN(*Att*)' is translated into

MIN($Att^C$)

b. MAX function.

The aggregate function MAX(·) returns the maximum attribute value that matches the query condition or that consists in a group. An aggregation operation like 'MAX(*Att*)' is translated into

$$\text{MAX}(Att^C)$$

c. COUNT function.

The aggregate function COUNT(·) returns the number of the tuples that match the query condition or that consist in a group. An aggregation operation like 'COUNT(*Att*)' is translated into

$$\text{COUNT}(Att^C)$$

d. SUM function.

The aggregate function SUM(·) returns the sum of an attribute that matches the query condition or that consists in a group. For an aggregation operation like 'SUM(*Att*)', its translation '*E*(SUM (*Att*))' is given by a secure two-party computation protocol.

Now recall that $\text{SUM}(E(Att))\leq E(\text{SUM}(Att))$ and $\text{SUM}(E'(Att))\geq E'(\text{SUM}(Att))$, so we have that $\textbf{Dec}(\text{SUM}(E(Att)), \boldsymbol{L})\leq \text{SUM}(Att)\leq \textbf{Dec}(\text{SUM}(E'(Att)), \boldsymbol{U'})$. See Protocol 1 for more details about the secure sum calculation protocol.

---

**Protocol 1** Secure sum calculation

**Input:** The server has *E*(*Att*) and *E'*(*Att*), such that their boundaries are generated by **SBoundaryGen**($\boldsymbol{R}$, $\sigma$) and **SBoundaryGen'**($\boldsymbol{R}$, $\sigma$) respectively.

**Output:** The client obtains SUM(*Att*).
The server obtains *E*(SUM(*Att*)).

**Protocol steps:**

1: The server requests SUM(*E*(*Att*)) and SUM(*E'*(*Att*)), and sends them to the client.

2: The client decrypts received values to get $d$=**Dec**(SUM(*E* (*Att*)), $\boldsymbol{L}$) and $d'$=**Dec**(SUM(*E'*(*Att*)), $\boldsymbol{U'}$).

3: If there is $d=d'$, then the client obtains SUM(*Att*)=$d$ and sends *E*(SUM(*Att*))=SUM(*E*(*Att*)) to the server.
This is the termination of this protocol.
Otherwise, continue to Step 4.

4: The client tries to find out a parameter $i$ ($0\leq i\leq d'-d$), s.t., **SumEqualityCom**(SUM(*E*(*Att*)), SUM(*E'*(*Att*)), $L[d+i]$, $U'[d+i]$) =0, which implies that SUM(*Att*)=$d+i$. Then the client sends $E(d+i)$ to the server.
This is the termination of this protocol.

---

Different from MIN, MAX and COUNT function, the nature of the OPEA algorithm determines that there are different approaches to invoke SUM function:

If SUM function exists in the select list of SELECT statement, then it will be translated into 'SUM($Att^C$), SUM($Att^C$_*Extension*)', where $*^C$_*Extension* represents the ciphertext column encrypted by the extended OPEA. If SUM function exists in the query condition and we further assume that the other operand is a constant, then it will be translated directly by **SumEqualityCom**(·).

e. AVG function.

The aggregate function AVG(·) returns the average of an attribute that matches the query condition or that consists in a group. An aggregation operation like 'AVG(*Att*)' is translated into

$$E(\text{SUM}(Att))/\text{COUNT}(Att^C)$$

where *E*(SUM(*Att*)) and COUNT($Att^C$) have been defined above. The way to invoke is similar to SUM function.

### 5.3.2 *Insert, Update and Delete*

Next, we publish several translation rules for the *insert*, *update* and *delete* manipulations.

1) Insert.

An insert manipulation is used to add new tuples to an existing table. The basic grammatical structure of INSERT statement is:

INSERT INTO <*table_name*> (<$Att_1$, $Att_2$, …>) VALUES ($val_1$, $val_2$, …)

In the encrypted database, the above statement is translated into

INSERT INTO <$table_name^C$> (<$Att_1^C$, $Att_2^C$, …>) VALUES ($E(val_1)$, $E(val_2)$, …)

INSERT statement supports the subquery, that is, it inserts the results returned by SELECT statement into the current table. See Section 5.3.1 for translation rules of the subquery.

2) Update.

An update manipulation is used to replace attribute values without changing all the other data in tuples. The basic grammatical structure of UPDATE statement is:

UPDATE <*table_name*> SET <*Att*>=*val*

In the encrypted database, the above statement is translated into

UPDATE <$table_name^C$> SET <$Att^C$>=$E(val)$

The FROM clause and the WHERE clause may also be used to restrict the scope of the tuples to be updated. And its update conditions are the same as the query conditions in SELECT.

3) Delete.

A delete manipulation is used to delete some tuples from a relation. The basic grammatical structure of DELETE statement is:

DELETE FROM <*table_name*>

In the encrypted database, the above statement is translated into

DELETE FROM <$table_name^C$>

The WHERE clause may also be used to restrict the scope of the tuples to be deleted. And its delete conditions are the same as the query conditions in SELECT.

## 5.4 Translation Rule of Relational Operation

Database relational operations can be divided into traditional operations and special operations. The former includes union, intersection, set difference, extended Cartesian product, etc., and the later includes selection, projection, join, etc. The following subsection expands on the translation rules of relational operations, which are used to translate the database relational operations over plaintexts into the corresponding operations over ciphertexts.

### *5.4.1 Traditional Relational Operation*

Since relations in database are essentially sets, traditional set operations can be performed on them. To keep things simple, we denote the relation with $n_1$ attributes (resp. $n_2$ attributes) as $P$ (resp. $Q$) in the plaintext database.

1) Union operator ($\bigcup$).

If $n_1$=$n_2$, then the union operator is implemented directly on their encrypted relations as below:

$$P^C \bigcup Q^C = \{p \mid p \in P^C \vee p \in Q^C\}$$

where duplicative tuples are eliminated by **EqualityCom**(·).

2) Intersection operator ($\bigcap$).

If $n_1$=$n_2$, then the intersection operator is implemented directly on their encrypted relations as below:

$$P^C \bigcap Q^C = \{p \mid p \in P^C, \exists q \in Q^C, \text{s.t.}, \mathbf{EqualityCom}(x,\ p,\ q) = 0\}$$

3) Set difference operator (-).

If $n_1$=$n_2$, then the set difference operator is implemented directly on their encrypted relations as below:

$$P^C\text{-}Q^C = \{p \mid p \in P^C, \forall q \in Q^C, \text{s.t.}, \mathbf{EqualityCom}(x,\ p,\ q) \neq 0\}$$

4) Extended Cartesian product operator (×).

An extended Cartesian product operator is implemented directly on their encrypted relations as below:

$$P^C \times Q^C = \{pq \mid p \in P^C \wedge q \in Q^C\}$$

where $pq$ represents a new tuple that is constituted by encrypted tuples $p$ and $q$ orderly.

### *5.4.2 Special Relational Operation*

Except the traditional set operations that are discussed above, there are some other special relational operations.

1) Selection operator (Δ).

In the selection operation, tuples are selected by given conditions to constitute a new relation. The translated selection operation is defined on an encrypted relation, that is,

$$\Delta_{Con}(P^C) = \{p \mid p \in P^C \wedge Con(p) = \text{'true'}\}$$

where $Con$ is a translated query condition over ciphertexts. If the condition holds for the encrypted tuple $p$, then it returns true.

2) Projection operator (Π).

In the projection operation, several attributes are selected from an existing relation to constitute a new relation. The translated projection operation is defined on an encrypted relation, that is,

$$\Pi_A(P^C) = \{p.A \mid p \in P^C\}$$

where $A$ is an encrypted attribute column or a set of encrypted attribute values, and $p.A$ is the corresponding component in the encrypted tuple $p$.

3) Join operator (⋈).

$P$ and $Q$ are connected together by a join operation to constitute a new data relation. There are many kinds of join operations, such as Equi-Join, $\theta$ Join and Natural Join. Taking $\theta$ Join as an example, the translated join operation is defined on encrypted relations, that is,

$$P^C \bowtie_{A\theta B} Q^C = \{p \wedge q \mid p \in P^C \wedge q \in Q^C \wedge p.A\ \theta\ p.B\}$$

where $\theta$ is a comparison operator (see the first part in Section 5.3.1). $A$ and $B$ are two encrypted attribute columns with the same semantics in the encrypted relation $P^C$ and $Q^C$, respectively.

To sum up, on the premise that no decryption is needed by SQL-Translator in T-DB, almost all kinds of data manipulations and relational operations on an encrypted database can be settled by translating SQL statements.

## 5.5 Functionality Limitation

We claim in this paper that T-DB is a nearly fully functional encrypted database system. What's more, all of its supported functionalities are obviously concentrated on common standard query types (the six kinds of clauses described in Section 5.3.1 are also regarded by Microsoft support document [36] as the most common SQL clauses). From the perspective of the syntax rules, the scope of the plaintext SQL statements that can be interpreted by SQL-Translator is only a subset of Transact-SQL standard syntax diagram. To this end, we analyze in this subsection the limitations in functionality of the T-DB translation module, which are mainly classified into the following several situations.

1) Encryption-Independent SQL statement.

Besides the DML statement, there is Data Definition Language (DDL) and Data Control Language (DCL) in Transact-SQL. The former is utilized to create, alter or drop database objects and their structural constraints, while the latter is utilized to grant or modify permissions for users and roles. SQL-Translator supports theoretically the translation of DDL and DCL statements. The translation process is equivalent to the anonymization of database entities.

For example, suppose that we have a plaintext DDL statement to create a data table:

CREATE TABLE *tbl* (*Att* int)

Only a simple translation is needed to get the corresponding encrypted DDL statement:

CREATE TABLE $tbl^C$ ($Att^C$ int)

No complex translation is involved in similar statements (such as ALTER, DROP, GRANT, DENY, REVOKE, and so on) when running on the encrypted database. They are entirely independent of the SQL-oriented encryption algorithm, and could be naturally achieved by any conventional anonymization technique. In addition, some auxiliary language elements like DECLARE, FETCH, OPEN and CLOSE directly apply to the encrypted database without statement translation.

2) UDF-Restricted SQL statement.

According to the translation rules described above, the equality comparison in the ciphertext domain has to invoke the two UDFs in Section 5.1. Both of them are performed as the computation or comparison on data records line by line. The flexibility of various equality comparisons in SQL statements is limited by the mathematical characteristics of our UDF, which cannot be settled just through the unmodified cloud database server.

Particularly, the **SumEqualityCom** function is more suitable for judgment on the order relation between an attribute column and a constant, instead of the order relation between two attribute columns. For example, a SQL clause like 'HAVING SUM($Att_1$)>SUM($Att_2$)' cannot be properly carried out before being decrypted by the database owner. Similarly, the DISTINCT keyword is designed to filter duplicative data records in the returned result set, so the **EqualityCom** function should be called many times. Nevertheless, in view of the row-based access in the cloud RDBMS, some methods (e.g., cursor) are necessary to be adopted when this function is operated on the same column. Thus, a huge workload in querying is inevitably involved in the encrypted database. Other important situations also appear in the ORDER BY clause and the GROUP BY clause, as detailed in Section 5.3.1.

3) Encryption-Restricted SQL statement.

It is quite hard for SQL-Translator to translate special SQL operators or keywords, that is, the following data manipulations are not contained in syntax rules due to encryption:

a. Arithmetical expression. Since it is often composed of other operations beyond addition, the arithmetical expression cannot be interpreted into the form of UDF. Compared with the outsourced scheme using homomorphic encryption techniques, OPEA does not have any distinct advantage in the numerical calculation over ciphertexts.

b. Built-in function. It is provided by Transact-SQL, such as the scalar mathematical functions (e.g., POWER, LOG, SIN, etc.), the aggregate functions (e.g., STDEV, STDEVP, VAR, VARP, etc.), and so on.

At present, the diagram of Transact-SQL syntax is growing and being improved. Enhanced DML clauses and options are emerging. More extended functionalities in the future may be done on the basis of our existing works.

In summary, T-DB system has already transcended all previous solutions in query functionality. Objectively, although it does not cover all the functionalities in Transact-SQL, it is enough to provide the basic form of ciphertext manipulations (as with the basic SQL functionalities generally required by Microsoft technology forum [37]). Therefore, T-DB has strong practicability.

SQL-Translator can translate any DML statement that meets the syntax rules, whose details are published in the online supplementary material (download at [35]) and Section 5.6. Among them, a more intuitive syntax comparison can be made to further analyze the baseset or limitation for functionality.

## 5.6 DML Syntax Rules (for T-DB)

Our SQL-Translator can translate any SQL statement that meets the following DML syntax rules. Errors or unexpected query results will be returned when these syntax rules are broken.

The SQL statement that is submitted by a database user should also pass the semantic and permission check in standard Transact-SQL.

The syntax rules are built using Transact-SQL syntax and arguments convention. Please see https://msdn.microsoft.com/en-us/library/ms177563.aspx for more details.

1) Syntax.

```
<SELECT statement> ::=
{
        SELECT <select_list>
        FROM <table_source>
        [WHERE <search_condition>]
        [<GROUP BY>]
        [HAVING <having_condition>]
        [ORDER BY {[table_name.]column_name [ASC | DESC]} [,…n]]
;}
```

<select_list> ::=
{[*table_name*.]* | [*table_name*.]*column_name* | <method_name> (<expression>)} [,…*n*]

<method_name> ::= {MIN | MAX | COUNT | SUM | AVG}

<expression> ::=
{[*table_name*.]*column_name* | *positive_integer*}

<table_source> ::=
{*table_name* | <joined_table>} [,...*n*]

<joined_table> ::=
{*table_name* <join_type> *table_name* ON <search_condition>}

<join_type> ::= {[INNER | LEFT | RIGHT | FULL] JOIN}

<search_condition> ::=
{
{<predicate> | (<search_condition>)}
[{ AND | OR } {<predicate> | (<search_condition>)}] [,…*n*]
}

<predicate> ::=
{
<expression> <comparison_operator> <expression>
| <expression> [NOT] BETWEEN <expression> AND <expression>
| <expression> IS [NOT] NULL
| <expression> [NOT] IN (<expression> [,…*n*])
| *table_name.column_name* [NOT] LIKE *pattern* [ESCAPE *escape_character*]
| <expression> <comparison_operator> (*restricted_subquery*)
| (*restricted_subquery*) <comparison_operator> <expression>
| [NOT] EXISTS (<subquery>)
}

<comparison_operator> ::= {<> | != | !> | <= | !< | >= | = | > | <}

<subquery> ::=
{
SELECT <select_list>
FROM <table_source>
[WHERE <search_condition>]
[<GROUP BY>]
[HAVING <having_condition>]
}

<GROUP BY> ::=
{GROUP BY [ALL] {*table_name.column_name*} [,…*n*]}

```
<having_condition> ::=
{
    {<predicate> | <having_predicate> | (<having_condition>)}
    [{ AND | OR } {<predicate> | <having_predicate> | (<having_condition>)}] [,…n]
}
```

```
<having_predicate> ::=
{
    <method_name> (<expression>) IS [NOT] NULL
    | <having_expression> <comparison_operator> <having_expression>
    | <having_expression> [NOT] BETWEEN <having_expression> AND <having_expression>
    | <having_expression> [NOT] IN (<having_expression> [,…n])
    | table_name.column_name [NOT] LIKE pattern [ESCAPE escape_character]
    | {SUM | AVG} (<expression>) <comparison_operator> positive_integer
    | {SUM | AVG} (<expression>) [NOT] BETWEEN positive_integer AND positive_integer
    | {SUM | AVG} (<expression>) [NOT] IN (positive_integer [,…n])
}
```

```
<having_expression> ::=
    {<expression> | {MIN | MAX | COUNT} (<expression>)}
```

```
<INSERT statement> ::=
{
    INSERT [INTO] {table_name} [(column_name [,...n])]
    {VALUES ({positive_integer | DEFAULT | NULL} [,...n]) [,...n] | <SELECT statement>}
;}
```

```
<UPDATE statement> ::=
{
    UPDATE {table_name} SET {column_name = {<expression> | DEFAULT | NULL}} [,...n]
    [FROM <table_source>]
    [WHERE <search_condition>]
;}
```

```
<DELETE statement> ::=
{
    DELETE [FROM] {table_name} [WHERE <search_condition>]
;}
```

2) Arguments.

*table_name*, *column_name* or *pattern*

cannot include any SQL keywords or the string 'subquery'.

*pattern*

can include the following valid wildcard characters: %, _, [] and [^].

*restricted_subquery*

can be considered a restricted <subquery> statement, where only a single value is returned by the subquery. SUM and AVG are not allowed in <method_name>.

## 6. SYSTEM PERFORMANCE

T-DB, whose performance is affected by various factors, can be deployed on any local or cloud RDBMS that supports UDF. This section provides some experimental results to evaluate T-DB from the following four aspects: (i) computational efficiency of OPEA; (ii) computational efficiency of SQL-Translator; (iii) functionality, precision, overhead and security in ciphertext manipulations; and (iv) the realistic database performance in Microsoft Windows Azure SQL Database (Microsoft SQL Azure for short).

## 6.1 Experimental Design

In order to avoid possible security risks from excessive user permissions, commercial cloud database products generally fail to support or open the programmable interface. In the latest version of SQL Azure, the user-defined function has been partially supported, but it still cannot deal with the storage or relation of encrypted data, which limits the application of certain outsourced encrypted database solutions.

In fact, there are several practical restrictions on the commercial SQL Azure platform, such as cloud migration, data type, data size, storage format, manipulation type, storage procedure, query execution time, etc. For the secure outsourcing schemes that rely on CLR components (e.g., CryptDB [22]), after its encryption and before its decryption, there exists a problem about the code conversion between character stream and byte stream. This problem is waiting for future systematic adjustment.

The execution efficiency of ciphertext SQL statements on the cloud is related with the cloud resources and server configuration, which is not the main concern of this paper. Hence, several simulation experiments are firstly given to make comprehensive evaluations about the overall performance of T-DB.

Experiments instantiated the database platform with Microsoft SQL Server 2012 and Transact-SQL language. We used two machines with 3.6GHz Intel Core i7 processor, 20GB memory and Windows 7 operating system to run the cloud client and the cloud server. OPEA and SQL-Translator modules were developed in the Microsoft Visual Studio 2015 Professional with C++11 standard and ran on the side of database owner. Cryptographic algorithms and the numerical calculation with arbitrary precision were implemented using the static library OPENSSL 1.0.1s [38] and boost 1.60.0 [39]. The T-DB implementation consists of approximately 2000 lines of source codes, mostly in the code for the encryption, decryption and interpreter module. The cloud server requires only approximately 30 lines of SQL codes to declare the UDFs that are used in executing ciphertext query manipulations.

The test database comes from TPC-H 2.17.1 [40], which is a decision support benchmark defined by TPC. It is a normalized and open database with controllable data size. TPC-H can simulate real-world business data. Its database diagram is presented in Figure 6. We set the TPC-H scale parameter to 10 (SF=10), then about 10GB relational tables and data items are generated. There are in total 86586082 records in eight tables.

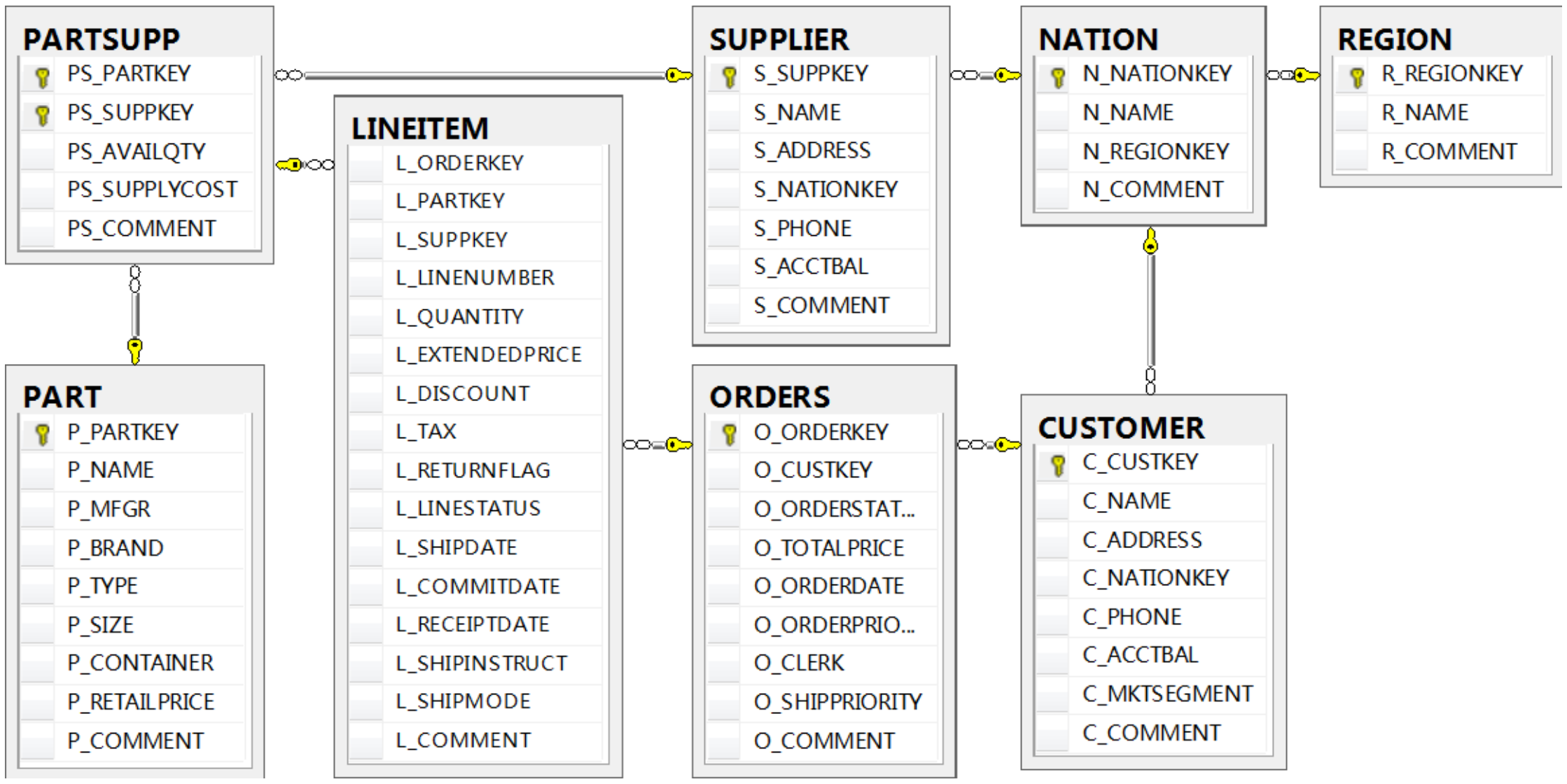

**Figure 6: Database diagram of TPC-H.**

The number of data records in each data table is listed in Table 1. TPC-H can well reflect the ability of a system to execute queries. It has complex data structure and various query types, so it is suitable for testing the effectiveness in executing SQL. TPC-H also makes the verification of OPEA and SQL-Translator comprehensive and objective.

**Table 1: Table size of TPC-H database.**

| Table name | Number of records | Table name | Number of records |
|---|---|---|---|
| Customer | 1500000 | Part | 2000000 |
| Lineitem | 59986052 | Partsupp | 8000000 |
| Nation | 25 | Region | 5 |
| Orders | 15000000 | Supplier | 100000 |

Below is a series of comparison tests on the T-DB functionality, which is the most valuable contribution. Their major target is to verify the practical usefulness of certain queries that are developed from the comparability of OPEA ciphertext. Here we introduce several selective SQL test cases (see Figure 7), including *query* (Q1 to Q5), *insert* (I6), *update* (U7) and *delete* (D8) statements. They cover nearly all standard query types in DML to demonstrate the improvement of our T-DB system on the query functionality of encrypted databases. The query conditions involved in Q1 to Q5 would be obviously impacted by the efficiency issues of core ciphers during their executions, which is particularly problematic in this environment. I6 to D8 are chosen to validate effective dynamic updates on ciphertexts. Take the complicated query Q4 for example: it tests comparison operators, GROUP BY clause and HAVING clause over multiple columns in an encrypted database simultaneously. This kind of query cannot be easily implemented with traditional cipher tools. Besides, these experiments could be extended to more test cases, and the experimental data in the paper is measured by taking the average time over at least ten iterations.

**Q1: Equality Query (=; Equi-Join; IS; IN; GROUP BY)**
```
SELECT S_SUPPKEY, S_NATIONKEY, N_REGIONKEY FROM
SUPPLIER JOIN NATION ON SUPPLIER.S_NATIONKEY=NATION.N_NATIONKEY;
```
**Q2: Range Query (>; <; ORDER BY; BETWEEN AND)**
```
SELECT C_CUSTKEY FROM CUSTOMER WHERE
C_NATIONKEY BETWEEN num1 AND num2 ORDER BY C_NATIONKEY;
```
**Q3: Aggregation Query (1) (MAX; MIN; COUNT)**
```
SELECT MIN(L_QUANTITY), MAX(L_QUANTITY), COUNT(L_QUANTITY) FROM
LINEITEM WHERE L_SUPPKEY=num1 AND L_LINENUMBER=num2;
```
**Q4: Aggregation Query (2) (SUM; AVG)**
```
SELECT PS_PARTKEY FROM PARTSUPP WHERE PS_PARTKEY<num1 GROUP BY
PARTSUPP.PS_PARTKEY HAVING SUM(PS_AVAILQTY)>num2;
```
**Q5: Fuzzy Query (Search; LIKE)**
```
SELECT O_ORDERSTATUS, O_ORDERPRIORITY FROM ORDERS WHERE
O_ORDERKEY<=num1 AND ORDERS.O_ORDERPRIORITY LIKE '%num2';
```
**I6: Insert**
```
INSERT INTO REGION(R_REGIONKEY, R_NAME) VALUES(num1, num2);
```
**U7: Update**
```
UPDATE REGION SET R_NAME=num2 WHERE R_REGIONKEY=num1;
```
**D8: Delete**
```
DELETE FROM REGION WHERE R_REGIONKEY=num1;
```

**Figure 7: Test cases of plaintext SQL statements.** Two random numbers (num1 and num2) are used here to return a non-empty result set.

## 6.2 Control Group

The experimental environment was built based on the system architecture in Section 3 to confirm the above conclusions. It strictly follows the procedures of secure database outsourcing. The cloud RDBMS server takes charge of managing encrypted databases and executing all queries received. Other works in the stage of pre-processing or post-processing are completed by the database owner.

The simulation took CryptDB [22] with default parameter setting (see Table 2) as the control group. Secret keys with the same length were selected to encrypt all columns. In CryptDB, data encryption and data decryption are performed by a trusted third-party proxy. During its pre-processing, some related Initial Vectors (IVs) are randomly generated and an AVL-tree is balanced by the range of plaintext values. The experimental parameters are listed in Table 2 where SEARCH algorithm [41] only leaks the number of search keywords, and the rate of false positives in searching is set to $1/2^{32}$.

**Table 2: Parameter setting of CryptDB.**

| Cipher algorithm | Parameter setting |
|---|---|
| AES-CBC | 16 bytes key; 16 bytes block |
| AES-CMC | 16 bytes key; 16 bytes block |
| Blowfish | 16 bytes key; 8 bytes block |
| OPE-AVL | 16 bytes key; 16 bits plaintext; 32 bits ciphertext |
| Paillier | 16 bytes key; 1024 bits parameter $n$; 256 bits parameter $a$ |
| JOIN-ADJ | 16 bytes key; 16 bytes block |
| SEARCH | 16 bytes key; 16 bytes ciphertext; 8 bytes salt |

Similarly, the OPEA algorithm in T-DB chose $T$+1 8-bit random numbers as its encryption keys. In other words, the total key length has reached 8($T$+1) bits, where $T$ is in the magnitude of $10^6$ for test data. The encryption costs of OPEA have nothing to do with the partition lengths, and only the storage efficiency of ciphertexts would make an effect on system performance, which is not primary measurement for T-DB. Other security parameters (e.g., symmetric keys or random salt values, etc.) are strong enough, so the comparison experiments that run against CryptDB may be considered fair in practice.

We conduct some theoretical comparisons with CryptDB (Table 3 and Table 4). The overwhelming advantage of T-DB is that it can be straightforwardly deployed without disturbing the CSP or database users. CryptDB has different levels of security, and is more secure than T-DB in some levels. However, after the outer onion decryptions of CryptDB are executed on the CSP for processing queries, it is also clear that T-DB will be at least as secure as CryptDB. Regarding the query efficiency and query precision on the database side, T-DB significantly outperforms CryptDB. And in terms of functionality, T-DB supports almost all types of efficient data manipulations (e.g., data query or data update) on encrypted databases. CryptDB in a fuzzy query only supports the full-word keyword search, excluding any form of regular expressions or the wildcard-based fuzzy queries.

**Table 3: Comparison on security, precision and efficiency.**

| Performance index | Experimental scheme | |
|---|---|---|
| | **T-DB** | **CryptDB** |
| Security (for OPE) | IND-AOCPA-secure | IND-OCPA-secure |
| Precision | Exact | False positive |
| Efficiency | High | Low |

**Table 4: Comparison on manipulation functionality.** Symbol √ and ⊗ indicate that the data manipulation is supported and partially supported by the experimental scheme, respectively.

| Manipulation functionality | Experimental scheme | |
|---|---|---|
| | T-DB | CryptDB |
| Equality query (e.g., =; IN; GROUP BY; Equi-Join; IS) | √ | √ |
| Range query (e.g., >; <; BETWEEN AND; ORDER BY) | √ | √ |
| Aggregation query I (e.g., MIN; MAX; COUNT) | √ | √ |
| Aggregation query II (e.g., SUM; AVG) | √ | ⊗ |
| Fuzzy query (e.g., Search; LIKE) | √ | ⊗ |
| Insert, delete, update | √ | √ |

Furthermore, in CryptDB a single plaintext column will always be transformed into multiple ciphertext columns to support queries on ciphertexts; in contrast, our T-DB produces one ciphertext column from each plaintext column, thereby saving much storage space. Beyond that, since incompatible encryptions (including a deterministic encryption AES-CMC, a non-deterministic encryption AES-CBC, an order preserving encryption OPE-AVL [27], a homomorphic encryption Paillier, an Equi-Join algorithm JOIN-ADJ and a keyword search algorithm SEARCH [41]) are adopted by CryptDB, the aggregation query II cannot apply to the HAVING clause. For certain queries where both summation and comparison are operated on the same column, the data items should be rewritten by homomorphic encryption and re-encrypted by trusted proxy, which leads to a huge computational cost.

## 6.3 Evaluation for OPEA

A primary problem in the efficiency evaluation of OPEA is the generation time of private boundaries. In view of the fact that the storage space of the boundaries is about twice that of the plaintext domain, the total time needed to generate boundaries will be longer as the $T$ value increases.

When evaluating the performance of OPEA, we first encrypted $T$ consecutive positive integers, and measured the average time to generate boundaries. Related experimental results and curves are given in Figure 8. The boundary generation time of OPEA gradually increases in proportion to the $T$ value, while the unit time for the simplified OPEA to generate private boundaries is stable within 0.17μs when $T$ is located from 1 to 100,000. For T-DB system, this is a kind of one-off overhead. Private partition boundaries are stored locally and not required to be exchanged in database outsourcing or query processing. That is to say, they are reused many times after being generated. Excessive computational overhead can be handled with the help of a private cloud, while the **SBoundaryGen** function (see Algorithm 4 for more details) is practically usable.

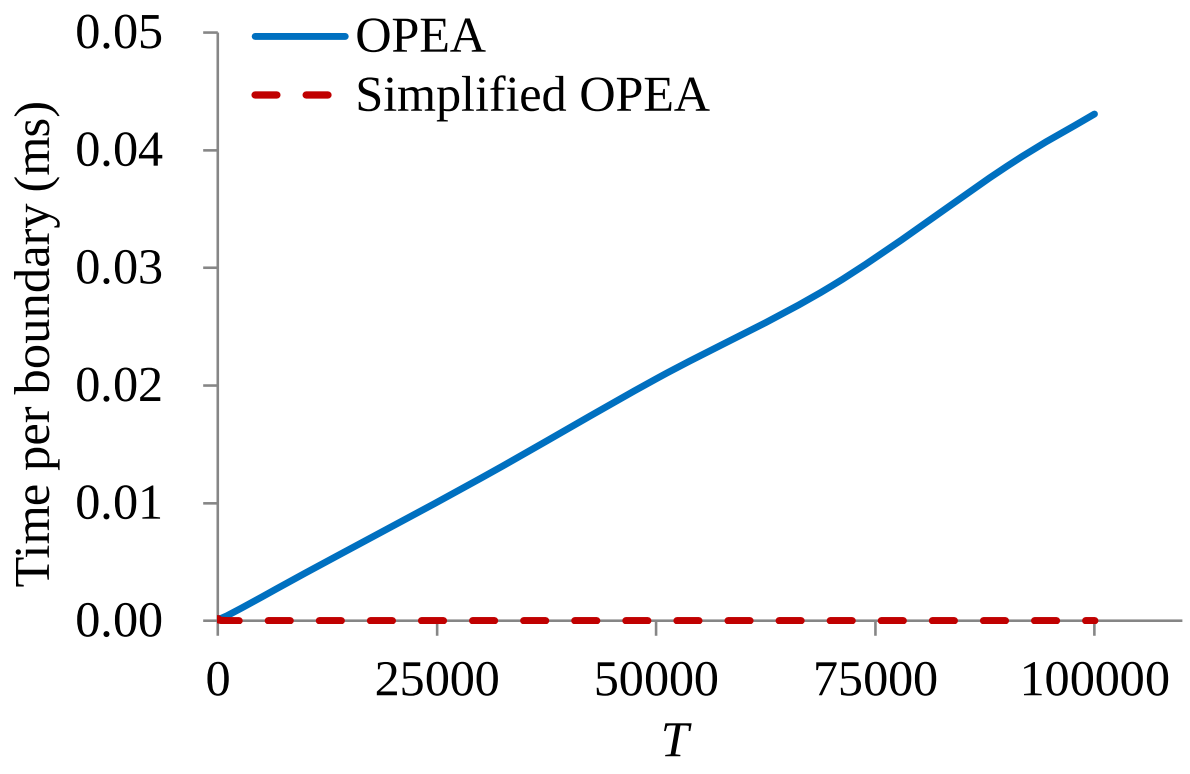

**Figure 8: Average time to generate boundary.**

After partition boundaries are generated, the encryption sub-algorithm of OPEA selects a random number from the corresponding partition and outputs it as a ciphertext, which can be finished in constant time. The decryption sub-algorithm searches and returns the index of the partition where the ciphertext is located, so an approximate logarithmic complexity can be achieved with the well-known binary search method.

Next, we separately performed encryption and decryption operations on the column *P_SIZE* of the table *Part*, and measured the average time and encryption frequency for each data item (Table 5). The encryption time of OPEA is about 1.74μs, second only to the symmetric cipher Blowfish. The decryption time of OPEA is the fastest (about 0.05μs), which indicates that the local computational cost is extremely low at the database owner side. The average comparison time at the cloud server side is nearly negligible (only about 0.03μs). Moreover, no additional network communication overhead is incurred as OPEA is non-interactive. The network traffic is just related with the size of the query request as well as the size of its result set.

**Table 5: Comparison on time-space overhead of cipher algorithms.** The measured average operation time in the third column includes the average time used in encryption/decryption, the conversion of storage type and the balance of AVL-tree (for OPE-AVL) on SQL Server. The encryption frequency in the last column refers to the total number of times encryption is required to make sure that the ciphertext column can support Equality Query, Range Query, Aggregation Query and Fuzzy Query simultaneously. Here, the outermost onion encryption is not taken into consideration. In SEARCH, we replace the data item with a string before encryption/decryption or matching test.

| Experimental scheme | Cipher algorithm | Average operation time (ms) | | | Encryption frequency |
|---|---|---|---|---|---|
| | | Encryption | Decryption | Other operation | |
| T-DB | OPEA | 0.00174 | 0.00005 | Compare: 0.00003 | 1 |
| CryptDB | AES-CBC | 0.01194 | 0.00644 | | 4 |
| | AES-CMC | 0.01204 | 0.00685 | | |
| | Blowfish | 0.00036 | 0.00029 | | |
| | OPE-AVL | 1.31980 | 1.31552 | Compare: 0.00001 | |
| | Paillier | 38.10740 | 2.51379 | Add: 0.06501 | |
| | JOIN-ADJ | 0.87301 | — | Adjust: 0.90835 | |
| | SEARCH | 0.01347 | 0.01219 | Match: 0.00757 | |

Also, the original data column is encrypted only once by T-DB, whereas each data column is encrypted 4 times by CryptDB, i.e., one column per allowable onion structure. Thus without consideration of the oversize ciphertexts, the storage space of the encrypted database in CryptDB is about 4 times the size of that in T-DB (i.e., the latter is almost of the size of the plaintext database).

## 6.4 Evaluation for SQL-Translator

The test cases in Figure 7 are translated by SQL-Translator into ciphertext SQL statements. We leave their translated results in the online material [35] and Figure 9. They can be submitted directly to the CSP to perform the same manipulation. The third column in Table 6 is the translation time. Evaluation results show that, under normal circumstances, the translation of DML statements can be done in microseconds.

```
Q1: Equality Query (=; Equi-Join; GROUP BY)
SELECT S_SUPPKEY^C, S_NATIONKEY^C, N_REGIONKEY^C FROM SUPPLIER^C
INNER JOIN NATION^C ON dbo.EqualityCom(x, SUPPLIER^C.S_NATIONKEY^C,
NATION^C.N_NATIONKEY^C) = 0;
Q2: Range Query (>; <; ORDER BY; BETWEEN AND)
SELECT C_CUSTKEY^C FROM CUSTOMER^C
WHERE (dbo.EqualityCom(x, C_NATIONKEY^C, num1^C) >= 0 AND
dbo.EqualityCom(x, C_NATIONKEY^C, num2^C) <= 0) ORDER BY C_NATIONKEY^C ASC;
Q3: Aggregation Query (1) (MAX; MIN; COUNT)
SELECT MIN(L_QUANTITY^C), MAX(L_QUANTITY^C), COUNT(L_QUANTITY^C)
FROM LINEITEM^C WHERE dbo.EqualityCom(x, L_SUPPKEY^C, num1^C) = 0 AND
dbo.EqualityCom(x, L_LINENUMBER^C, num2^C) = 0;
Q4: Aggregation Query (2) (SUM; AVG)
SELECT *, (SELECT TOP 1 PS_PARTKEY^C FROM PARTSUPP^C PARTSUPP^C_A WHERE
dbo.EqualityCom(100, PARTSUPP^C_A.PS_PARTKEY^C, PARTSUPP^C_B. PS_PARTKEY^C)
= 0) AS PS_PARTKEY^C_Group INTO #TEMPORARY_TABLE1 FROM PARTSUPP^C
PARTSUPP^C_B WHERE dbo.EqualityCom(100, PS_PARTKEY^C, num1^C) < 0;
SELECT PS_PARTKEY^C_Group FROM #TEMPORARY_TABLE1 GROUP BY
PS_PARTKEY^C_Group HAVING dbo.SumEqualityCom(SUM(PS_AVAILQTY^C),
SUM(PS_AVAILQTY^C_Extension), L(num2), U'(num2)) > 0;
DROP TABLE #TEMPORARY_TABLE1;
Q5: Fuzzy Query (Search; LIKE)
ALTER TABLE ORDERS^C ADD O_ORDERPRIORITY^C_Match2 BIT NOT NULL
CONSTRAINT O_ORDERPRIORITY^C_DF2 DEFAULT 0
DECLARE @StrCol varchar(8000), @IsMatch bit, @RowCount int
DECLARE @CharCol0 int, @CharCol1 int
DECLARE STRCUR CURSOR FOR SELECT O_ORDERPRIORITY^C FROM ORDERS^C
OPEN STRCUR
FETCH NEXT FROM STRCUR INTO @StrCol
WHILE @@FETCH_STATUS = 0 BEGIN
  SET @RowCount = 0; SET @IsMatch = 1
  DECLARE CHARCUR SCROLL CURSOR FOR SELECT * FROM dbo.Split(@StrCol, ';')
  OPEN CHARCUR
  FETCH LAST FROM CHARCUR INTO @CharCol0
  IF @@CURSOR_ROWS - 1 >= 1 AND @CharCol0 >= 0 BEGIN
    IF @@CURSOR_ROWS - 1 - @RowCount >= 1 BEGIN
      FETCH LAST FROM CHARCUR INTO @CharCol1
      FETCH PRIOR FROM CHARCUR INTO @CharCol1
      IF dbo.EqualityCom(x, @CharCol1, num2^C) != 0 SET @IsMatch = 0
    END
  END
  ELSE SET @IsMatch = 0
  CLOSE CHARCUR; DEALLOCATE CHARCUR
  IF @IsMatch = 1 UPDATE ORDERS^C SET O_ORDERPRIORITY^C_Match2 = 1 WHERE
CURRENT OF STRCUR
  FETCH NEXT FROM STRCUR INTO @StrCol
END
CLOSE STRCUR; DEALLOCATE STRCUR
SELECT O_ORDERSTATUS^C, O_ORDERPRIORITY^C FROM ORDERS^C
WHERE dbo.EqualityCom(x, O_ORDERKEY^C, num1^C) <= 0 AND
ORDERS^C.O_ORDERPRIORITY^C_Match2 = 1;
ALTER TABLE ORDERS^C DROP CONSTRAINT O_ORDERPRIORITY^C_DF2;
ALTER TABLE ORDERS^C DROP COLUMN O_ORDERPRIORITY^C_Match2;
I6: Insert
INSERT INTO REGION^C(R_REGIONKEY^C, R_NAME^C) VALUES (num1^C, num2^C) ;
U7: Update
UPDATE REGION^C SET R_NAME^C = num2^C
WHERE dbo.EqualityCom(x, R_REGIONKEY^C, num1^C) = 0;
D8: Delete
DELETE FROM REGION^C WHERE dbo.EqualityCom(x, R_REGIONKEY^C, num1^C) = 0;
```

**Figure 9: Translated ciphertext SQL statements.** Italic represents the table names or column names. $*^C$ and $*^C$*_Extension* indicate the corresponding encrypted value in the ciphertext domain of AES/OPEA and the extended OPEA, respectively.

The next translation experiments were studied for some nested test cases (Figure 10a) and unit test cases (Figure 10b). The former tests the average translation time of the nested queries in SELECT statement. It grows linearly with the increase of nesting levels. As for the latter, the test object of unit testing is the minimal query unit in a SELECT statement that contains several keywords or operators. Due to the complex translation rules of fuzzy queries, SQL-Translator spent much more time (about 0.19ms) to translate LIKE operator than other cases, but its overhead is still considerably low.

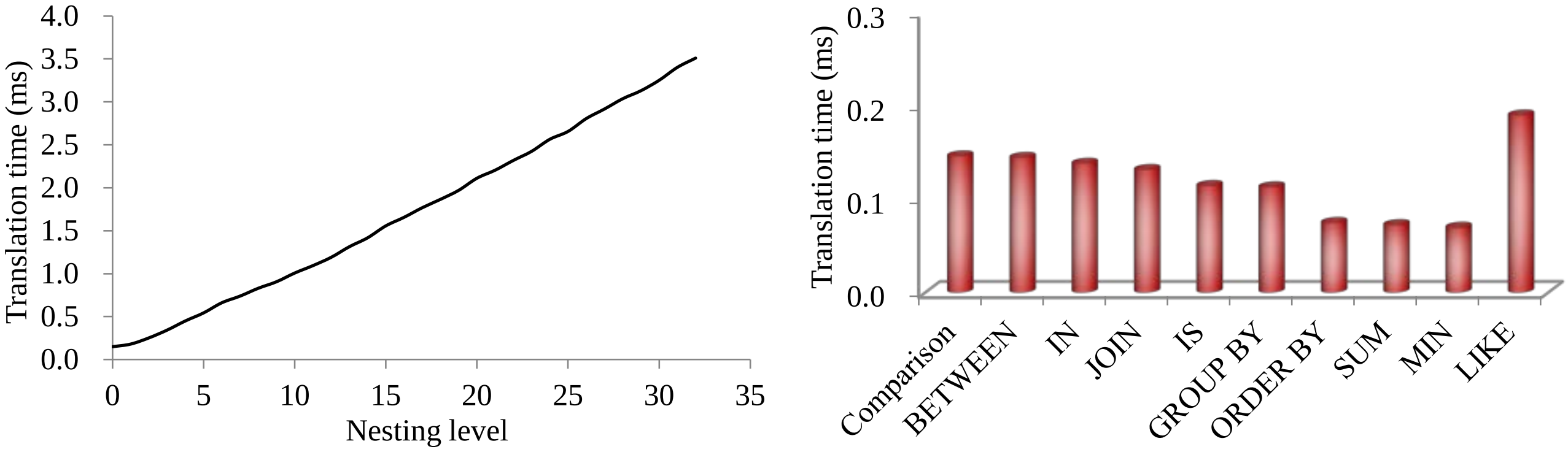

**(a) Nested testing** **(b) Unit testing**

**Figure 10: Translation time of SQL-Translator.**

To help us to further understand the translation principle of the subquery, we consider the following basic nested query statement:

SELECT * FROM $tbl_1$ WHERE $Att_1$!=(SELECT $Att_2$ FROM $tbl_2$ WHERE $Att_3$=*num*)

where $tbl_1$ and $tbl_2$ are table names, $Att_1$, $Att_2$ and $Att_3$ are column names, and *num* is a constant. The above plaintext statement will be translated by SQL-Translator into the following set of ciphertext query statements:

DECLARE @inter_var2 int;
SELECT @inter_var2=$Att_2^C$ FROM $tbl_2^C$ WHERE **EqualityCom**($x$, $Att_3^C$, $E(num)$)=0;
SELECT * FROM $tbl_1^C$ WHERE **EqualityCom**($x$, $Att_1^C$, @inter_var2)!=0;

These statements would actually run directly over encrypted databases to support nested queries on ciphertexts, where @inter_var2 is a corresponding temporary intermediate variable.

## 6.5 Evaluation for Ciphertext SQL

In T-DB, a data manipulation request from the database user will be sent to a cloud RDBMS after being translated by the database owner. The T-DB workload of query processing is transferred to the cloud as much as possible in order to maximize the query precision in the encrypted database. Since there is no false positive or false negative in the query results returned, the overhead of post-processing (to eliminate false positives) is minimized.

Since CryptDB does not follow DBaaS and cannot be directly deployed on SQL Azure, the subsequent comparisons were conducted in the local simulation environment with a simulating CSP works as required by CryptDB. The query processing time of CryptDB in the encrypted database consists of two parts, i.e., the time for onion decryption and column updates before query, and the time for execution of the ciphertext SQL statement. Overall, T-DB has better efficiency in query and decryption than CryptDB.

As for Q4, the ciphertext manipulation in CryptDB involves a range query and an aggregation query with additive homomorphism. Since the comparison and summation cannot be operated on the same column, several false positives were thus introduced. The overhead of a post-processing stage includes the time for pre-processing and decryption of OPE-AVL, the time for decryption of ciphertext addition as well as the time for elimination of false positives. The encrypted column *PS_AVAILQTY* was first decrypted by the database proxy, then its order relation was judged. This process took about 5840.245ms, i.e., 0.171 queries per second.

As for Q5, the ciphertext manipulation in CryptDB involves a range query and a keyword matching query. Since the keyword position cannot be matched by the wildcard characters in SEARCH algorithm, a multitude of false positives were introduced. The overhead of the post-processing stage includes the time used in decryption of AES-CBC and SEARCH, and the time used in elimination of false positives. This process took about 13011.340ms, i.e., 0.077 queries per second.

Regarding the post-processing efficiency at the database owner side as well as the query precision and the time overhead of the CSP, Figures 11a to 11c indicate that our T-DB has much higher query precision and far lower time overhead than CryptDB.

The specific analysis is shown as follows:

In Figure 11a, we investigate the average queries per second on the database owner, and then compare with a CryptDB proxy running on the same encrypted database. Without considering the communication overhead and query overhead of the cloud RDBMS server, it can be observed from this figure that T-DB decrypts the query results efficiently in the post-processing stage. Our throughput is at least 36.04 times higher than CryptDB, and is even up to $10^4$ times faster in some particular test cases.

In terms of the query precision of the CSP, a 100% recall rate is acquired in both T-DB and CryptDB, while their precision rate is quite different. A careful observation of Figure 11b indicates clearly that T-DB preserves the precision rate as a plaintext database in all of the five test cases, but the average precision of CryptDB in Q4 and Q5 (false positives) is 43% and 50%, respectively. This figure also reveals that the homomorphic encryption technique is not enough to achieve accurate ciphertext manipulations, especially in the range query and the fuzzy query. Take Q4 with Paillier homomorphic algorithm as an example, CryptDB cannot perform any ciphertext comparison before its decryption, so the false positives coming from the CSP are inevitable.

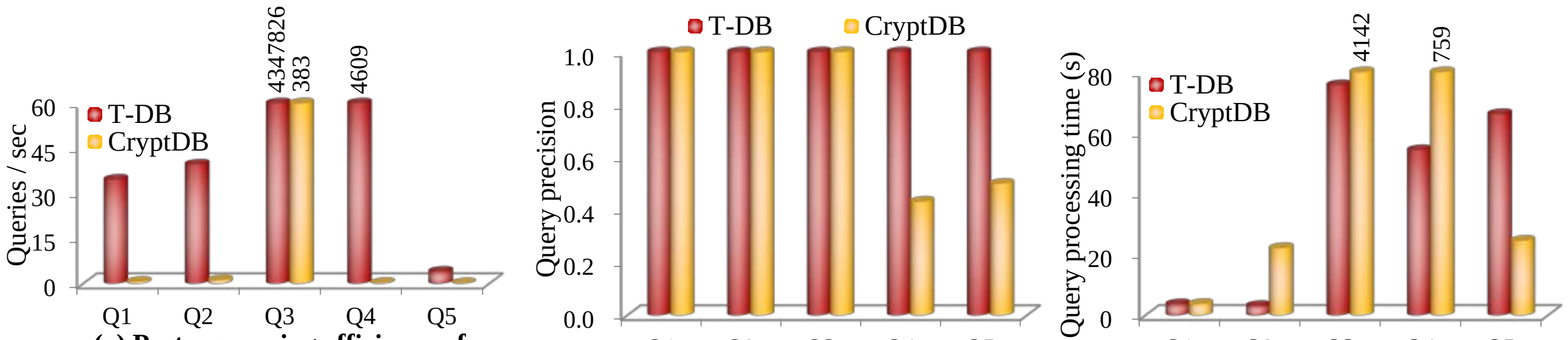

**(a) Post-processing efficiency of database owner**

**(b) Query precision of CSP**

**(c) Query processing time of CSP**

**Figure 11: Performance comparison between T-DB and CryptDB.**

Figure 11c compares the query processing time of the CSP. It is noted that T-DB can process the ciphertext queries more efficiently than CryptDB except the fuzzy query Q5. However, CryptDB may return a great number of erroneous results. Although the CSP takes less time to execute Q5, excessive post-processing overhead is transferred to the database owner. Besides, according to Table 5 and the comparison results of Q4 in this figure, we point out that T-DB consumes a shorter time than CryptDB either in the client-side encryption/decryption processing or in the server-side query processing. For this reason, it can be seen that the impact of the OPEA algorithm on the overall query efficiency is outweighed by the homomorphic encryption algorithm (e.g., Paillier).

In conclusion, compared with CryptDB, our T-DB has the following advantages:

1) T-DB can be straightforwardly deployed without disturbing the CSP or the database users.

2) T-DB can handle almost all kinds of basic DML statements, while CryptDB cannot.

3) The CSP does not participate in the decryption of outsourced data, thereby avoiding the leakage of secret keys.

4) Only a single cipher algorithm is adopted in T-DB. Thus the storage overhead of encrypted databases is lower.

5) The average query efficiency of the CSP is higher under the same query precision.

6) Some wildcard-based pattern matching is supported in fuzzy queries, which perfects its query functionality.

7) There is no need for database owners to eliminate the false positives. Thus the local amount of communication and computation is smaller.

8) If an aggregation query and a range query are operated on the same column, then the query precision is higher and the rate of false positives is lower in T-DB.

The source code and other test cases related to T-DB may be downloaded at [35]. Please see Section 7.6 for more detailed illustrations.

## 6.6 Evaluation for SQL Azure

To make a better evaluation of the applicability of T-DB on realistic cloud computing platforms, we implemented this system in SQL Azure environment. The client has exactly the same configuration as that in Section 6.1. The relational SQL database service with the level of Standard S3 was qualified through a free Azure account in China [42], and the server has 100 Database Transaction Units (DTUs). A DTU is a blended measure of CPU, memory and data I/O rate. The above two machines were connected over a shared 300Mbps wireless network. The client configuration ensures that neither the database users nor the database owner is the bottleneck in any case.

A third-party tool called SQLAzureMW v4x [43] was applied to help the database owner to migrate her existing databases to SQL Azure engine (the same test database and cases are adopted here, and please refer to previous sections for specific data size, data source and experimental parameters). It can solve a series of possible restrictions on compatibility during the migration (e.g., the lack of the clustered index, etc.). In the real experiment mentioned below, all of the data items, the UDF definitions and the database architecture in the encrypted TPC-H are stored by SQL Azure, and a fully managed database service is also provided. Beyond that, there appears to be no change on the relational structure of outsourced databases or the access mode of database users.

The total execution time of data manipulations on the encrypted database includes three parts, i.e., translation time of SQL statements, execution time of ciphertext SQL, and decryption time of the final query results. Generally, the query execution time of an encrypted database is greater than that of a plaintext database (see Table 6). Further analyses reveal that the overall performance of T-DB is severely restricted by the execution efficiency of ciphertext statements. The most frequent queries, Q1 to Q2, can be processed in seconds. The time overhead of the unusual queries, Q3 to Q5, is much greater. This performance difference is mainly due to the elastic computing mechanism on the cloud.

The database user is allowed to rent cloud resources as flexible as possible, depending on actual business requirements and usage conditions. Afterwards a dynamic resource scheduling algorithm is utilized by the CSP to achieve load balance. A large number of query requests that come from the database owner are allocated on demand to specific cloud servers. No artificial factor is involved when the queries are running on SQL Azure, but their execution time (see Table 6) might be influenced by the tuning strategies of SQL execution plan. Some queries, for instance, the aggregation operations in Q3, spend a longer amount of time in both plaintext database and encrypted database, which is somewhat different as that in the local simulation experiment. Therefore, it is simply too difficult to ascertain the main reason for causing the current statistics.

**Table 6: Comparison on execution time of data manipulations between plaintext database and encrypted database in a realistic cloud.** (*n*) in the rightmost column indicates that there are *n* components in the returned record. Statement execution time in the fourth column represents the total response time of SQL Azure, i.e., the average time elapsed between the time when a query request is submitted by the database owner and the time when its query results are received (end-to-end delay included).

| Test case | Execution time in plaintext database (ms) | Execution time in encrypted database (ms) | | | | Number of returned records |
|---|---|---|---|---|---|---|
| | | SQL translation | SQL execution | Result decryption | Total | |
| Q1 | 511.87 | 0.36088 | 20404.00 | 28.93420 | 20433.30 | 100000 (3) |
| Q2 | 368.27 | 0.18138 | 18629.53 | 25.22940 | 18654.94 | 119793 (1) |
| Q3 | 216387.33 | 0.18771 | 509621.73 | 0.00023 | 509621.92 | 1 (3) |
| Q4 | 3.47 | 0.27632 | 317658.60 | 0.21696 | 317659.09 | 863 (1) |
| Q5 | 780.93 | 0.28018 | 329690.00 | 253.06360 | 329943.34 | 50103 (2) |
| I6 | 2.20 | 0.08120 | 3.40 | 0 | 3.48 | 0 |
| U7 | 2.87 | 0.11712 | 3.13 | 0 | 3.25 | 0 |
| D8 | 3.07 | 0.10167 | 1.87 | 0 | 1.97 | 0 |

Additionally, the tendency of the total response time in the encrypted database under different performance levels, i.e., different number of DTUs, is shown as Figure 12. We just take Q1 and Q2 as representatives. The total response time (i.e., the query efficiency of T-DB) in SQL Azure gradually decreases with the increase of the number of DTUs.

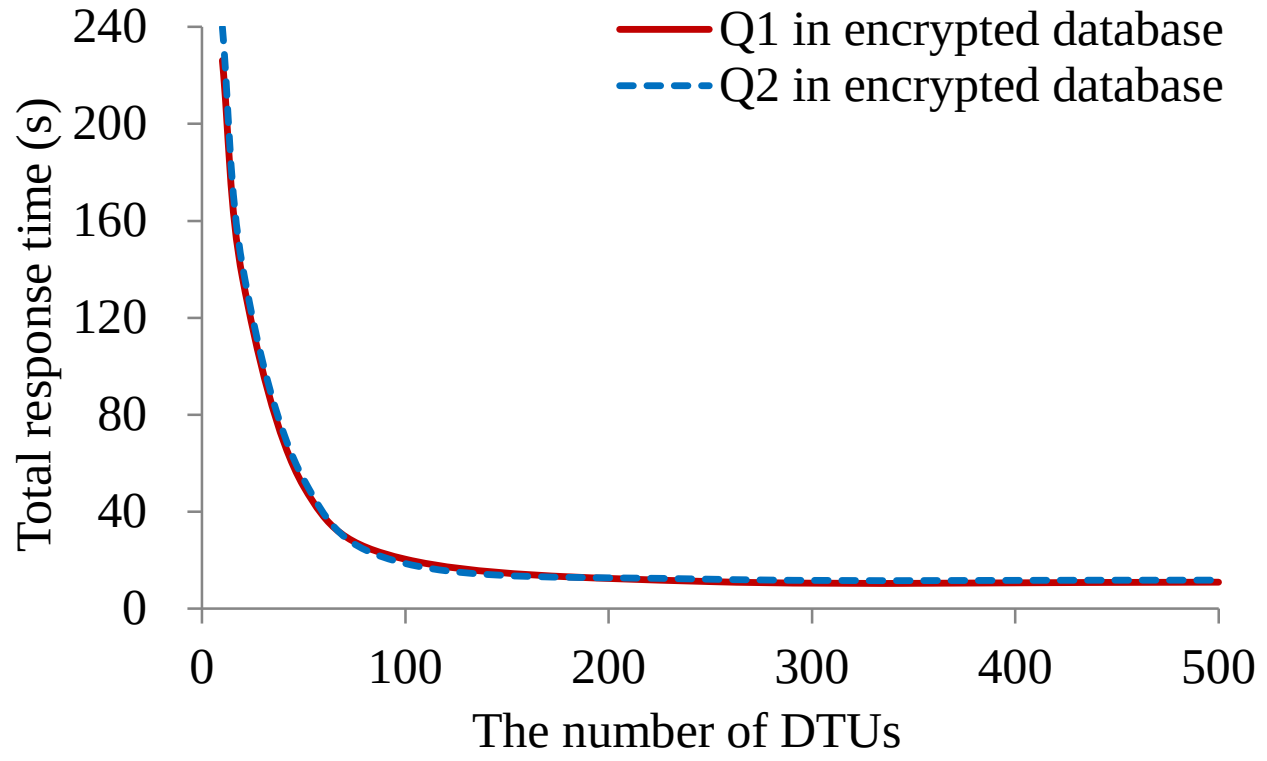

**Figure 12: Total response time in Microsoft Windows Azure SQL Database with different performance levels.**

The T-DB applicability in a realistic cloud environment has already been well validated. The migration to SQL Azure is available for T-DB.

# 7. DISCUSSION

To begin with, this section discusses the system security of T-DB, the noise growth in OPEA as well as the data type and data length in SQL-Translator. Here the influence of different encryption rules on the encrypted fields is analyzed. We also summarize the related works and point out the possible direction for future research.

## 7.1 System Security

The security under two threat models is analyzed on the system level. T-DB aims to provide the following security guarantees:

1) Security guarantee against Threat 1.

The database owner uses OPEA and a deterministic encryption algorithm to encrypt all sensitive outsourced data and names of tables or columns. Decryption keys are stored locally and isolated from the cloud, so it is quite difficult for the attacker to compromise any outsourced encrypted data by a ciphertext-only attack. Besides, T-DB does not hide sensitive side information, such as the overall table structure, the number of rows or columns, the field type, and the data size.

Some strict access control policies can be enforced by the database owner. Tables or columns with different security levels are encrypted with different keys to minimize the correlation leakage. Meanwhile, we regulate that an authorized user is allowed to log into the system by sharing a password. The database owner takes the access permissions of the currently logged-in user to lookup his encryption keys. Thus, the accessed private data can only be encrypted or decrypted correctly; the submitted SQL can only be translated properly, if that user is requesting his authorized contents. Otherwise, these manipulations will be poorly executed.

Executing ciphertext queries needs to encrypt the query conditions with different secret keys according to the specified request. The data relations revealed by executing SQL statements depend on the scope of query objects. For instance, if encryption granularity is an entire column, then the $x$ selected by SQL-Translator can only be used to run several specific data queries (i.e., one or more of these queries that are most commonly in equality, range, aggregation and fuzzy) on the current column. That is to say, the actual contents of data items and sensitive information of other columns will not be leaked.

The attacker at the cloud cannot translate the plaintext SQL. It is even harder for him to request any data manipulations on the CSP, so the visits to confidential data are minimized at the cloud.

However, since basic grammatical structure is preserved in the process of statement translation, a malicious attacker could still make a guess based on translated ciphertext SQL statements and steal related information about the private data, mainly including the table structure, entity relations, data item distributions, etc. Specifically, if the cloud receives an Equi-Join query request on two encrypted fields, i.e., $Att_1^C$ and $Att_2^C$, then the attacker can confirm that the data in these two columns are encrypted with the same key, and they have similar semantics as well as precise equivalence. For another example, the attacker can infer a general distribution of current attribute values according to the number of query results returned from the cloud server. What's more, by simply combining and concating with other ciphertext SQL clauses, the query scope will be gradually narrowed in order to detail the distribution of the encrypted database. But indeed the above attack strategies provide few helps for analyzing boundaries of ciphertext partitions.

2) Security guarantee against Threat 2.

Firstly, if a malicious database user colludes with the CSP, then a chosen-plaintext attack will be launched. Ciphertext queries can be performed adaptively by a cloud attacker in the database user's authorized access domain. Some relationships between plaintexts and ciphertexts are established from the final query results. Other private data encrypted with the same key in legitimate cloud servers may thus suffer high privacy risks. For instance, in the case that all of the secret keys and private boundaries are unknown, the malicious cloud attacker can construct an SQL statement that contains specified query conditions, consequently, the input and output of SQL-Translator can be controlled to request the encryption of sequential plaintext values in the form of a SQL injection attack. At last, in accordance with the ideas of order-preservation, additivity and clustering, a general interval and an approximate distribution of ciphertext partitions are analyzed by the attacker. Another fresh encrypted value will be classified into the closest partition with a high probability.

Collusion attack is one of the common security issues in DBaaS framework. In practical terms, every time we execute a data manipulation in an outsourced encrypted database, a certain number of correspondences between plaintexts and ciphertexts are established. Hence, any cipher algorithm with order-preservation cannot provide security guarantee under this attack, that is, collusion risk is inevitable in DBaaS, as T-DB does.

Secondly, in order to minimize the data leakage, some strategies about fine-grained access control and key management (e.g., [44, 45], etc.) have to be designed and implemented on the database owner. At any time, T-DB leaks at most the query requests or query results of the currently logged-in cloud users to the collusion attacker, while the private data not accessed would remain confidential since different data are encrypted with different secret keys. Considering that if the collusion attacker has limited access permission, the loss of sensitive information is still under control.

Moreover, the second threat model also includes some active attacks. A malicious database user who is involved in the collusion can insert or update the cloud data if he is granted the corresponding permissions, which may indirectly compromise secrecy. To prevent this kind of attack, we have to introduce appropriate mechanisms for integrity verification and automated data recovery.

In conclusion, our T-DB is a practical secure outsourcing system with the fundamental security on the premise of supporting almost fully functional data manipulations over encrypted databases.

## 7.2 Noise Growth

Essentially, a random noise is added into ciphertexts while encrypting in the OPEA algorithm. The selection of noise should meet some special conditions to ensure cryptographic security as well as order-preservation and additivity. Though the regular decryption function in OPEA can return the only and right plaintext value, serious noise growth always appears after additive operations. Since direct decrypting encrypted sum causes errors if the noise growth is greater than a certain threshold, this subsection analyzes the influence of noise growth that is caused by summation in the ciphertext domain on the correctness of decryption.

In short, the summation of ciphertexts in probabilistic OPEA algorithm may add noise to decryption and in the worst case, the decryption may fail. From the research in Section 4.6, we explain this phenomenon in three aspects.

1) The probability of the deviation of ciphertext partition.

The random noise grows after several ciphertexts are added together, and the probability of the deviation of ciphertext partition under various summation frequencies is different. Take the summation between two numbers as an example, for any positive integers $a$, $b$, $c=a+b$ ($a, b, c \in X$), let us set $d$=**Dec**($E(a)+E(b)$, $\boldsymbol{L}$) and $d'$=**Dec**($E'(a)+E'(b)$, $\boldsymbol{U'}$).

Firstly, from Theorem 2, we have

$$L[a]+L[b] \le E(a)+E(b) \le U[a]+U[b] \le L[a+b]=L[c]$$

So their encrypted sum $E(a)+E(b)$ is decrypted correctly only if the equation $E(a)+E(b)=U[a]+U[b]=L[a+b]$ holds, then it is true that

$$P[d=c] \le \frac{1}{U[a]+U[b]-(L[a]+L[b])} = \frac{1}{R_a+R_b}$$

And since the inequation $a+b>c-1$ always holds, a conclusion follows Theorem 6 (see Section 4.4.3), that is,

$$L[c-1] < E(a)+E(b) \le L[c]$$

where there is $E(a)$=RandomSelect($L[a]$, $L[a]+1$, $L[a]+2$, …, $U[a]$) and $E(b)$=RandomSelect($L[b]$, $L[b]+1$, $L[b]+2$, …, $U[b]$). The returned value of random selection function RandomSelect(·) is the noise in OPEA. If all of the random numbers involved in summation are small, then the decryption of the encrypted sum will go wrong, that is, the function **Dec**($E(a)+E(b)$, $\boldsymbol{L}$) may return $c$ or $c$-1, thus,

$$P[d=c-1] = 1-P[d=c] \ge 1-\frac{1}{R_a+R_b}$$

Evidenced by the similar token, from Theorem 4, we have

$$U'[c]=U'[a+b] \le L'[a]+L'[b] \le E'(a)+E'(b) \le U'[a]+U'[b]$$

So their encrypted sum $E'(a)+E'(b)$ is decrypted correctly only if the equation $U'[a+b]=L'[a]+L'[b]=E'(a)+E'(b)$ holds, then it is true that

$$P[d' = c] \le \frac{1}{U'[a] + U'[b] - (L'[a] + L'[b])} = \frac{1}{R_a + R_b}$$

And since the inequation $a+b<c+1$ always holds, a conclusion follows Theorem 7 (see Section 4.4.3), that is,

$$U'[c] \le E'(a)+E'(b) < U'[c+1]$$

The function **Dec**($E'(a)+E'(b)$, ***U'***) may return $c$ or $c+1$, thus,

$$P[d' = c+1] = 1 - P[d' = c] \ge 1 - \frac{1}{R_a + R_b}$$

In summary, whether the OPEA algorithm is used alone or the extended OPEA algorithm is used alone, the probability of the deviation of ciphertext partition caused by summation is at least $P[d=c-1]$ or $P[d'=c+1]$. It approaches one with the increase of $R_a$ and $R_b$. The above discussion indicates that summation increases the noise, and it is easy to see that multiplication also does so.

As for the next generalized case that *value*=SUM($\varsigma$), $|\varsigma|\ge 2$, we set $d$=**Dec**(SUM($E(\varsigma)$), ***L***) and $d'$=**Dec**(SUM($E'(\varsigma)$), ***U'***). From the theorems in Section 4.3 and Section 4.4, the upper probability that the encrypted sum is decrypted correctly is shown as follows:

$$P[d = value] = P[d' = value] \le \frac{1}{\sum_{t\in\varsigma} R_t}$$

And the probability of the deviation of ciphertext partition caused by summation is at least

$$P[d = value - 1] = P[d' = value + 1] \ge 1 - \frac{1}{\sum_{t\in\varsigma} R_t}$$

which approaches one with the increase of $R_t$ ($t\in\varsigma$).

2) Client-side overhead in secure sum calculation protocol.

With the help of secure sum calculation protocol, a judgment about the encrypted sum is made by the decrypted results from OPEA and its extended version. Since $d\in\{value-1, value\}$ and $d'\in\{value, value+1\}$, the probability that Step 3 in Protocol 1 outputs directly the correct plaintext sum by a one-time comparison is shown as follows:

$$P[d = d' = value] \le \left(\sum_{t\in\varsigma} R_t\right)^{-2}$$

In other cases, the correct SUM($\varsigma$) is located between $d$ and $d'$. If $d'-d=2$ at this time, then $value=d+1=d'-1$, otherwise the client should invoke the **SumEqualityCom**($\cdot$) function to get it. It is obvious that only a finite number of comparison operations are needed, so the computation complexity to eliminate decryption error is constant.

3) Generalized method to control noise growth.

Secure sum calculation protocol solves the problem about the decryption error which is caused by noise growth. As we note in the above derivation with further analysis of Theorem 6 and Theorem 7, the lower and upper limit of encrypted sum is on the premise that its system keys satisfy that $\sigma \ge (value+|\varsigma|)\cdot R_T - R_1$. This means that for a fixed $\sigma$, both the summation frequency $|\varsigma|$ and the plaintext sum *value* are limited, otherwise there will be a wrong decryption. In other words, we demonstrate that the noise is controllable and its growth can be effectively controlled by simply adjusting $\sigma$.

## 7.3 Data Type and Data Length

In general, the OPEA encryption function takes integers as input, but it also provides some extended encryption interfaces for float type or string type. SQL-Translator can easily translate data of arbitrary type and arbitrary length. Descriptions of storage format are given below.

For a float data, a simple approach is to divide a ciphertext domain by the least precision of plaintext domain. Lower and upper boundaries are recomputed on the basis of partition distribution. For a character data (e.g., a plaintext string '*abcd*'), it is split into multiple characters using standard delimiters (e.g. a semicolon) in CryptDB [22], and then each of the characters is encrypted by SEARCH algorithm [41]. In other words, the encrypted string is denoted as

$$E_{SEARCH}(a);\ E_{SEARCH}(b);\ E_{SEARCH}(c);\ E_{SEARCH}(d)$$

In T-DB, the database owner transforms it into numerical data (e.g., English characters should be transformed into ASCII code; Chinese characters should be transformed into Unicode) before encryption, following three encryption rules:

1) Field in wildcard-based fuzzy query.

For the fields in wildcard-based fuzzy queries, the plaintext string is transformed into ASCII code and then encrypted in sequence, so the encrypted string can be denoted as

$$E_{OPEA}((\text{int})a);\ E_{OPEA}((\text{int})b);\ E_{OPEA}((\text{int})c);\ E_{OPEA}((\text{int})d);\ \pi$$

where the minimum matching unit is assumed as a character. $\pi$ is the number of trailing blanks in the plaintext string, which is utilized to match a search pattern.

2) Field in equality, range and aggregation query.

For the fields in the equality, range and aggregation queries, the whole plaintext string can be transformed into numerical data and then encrypted, so the encrypted string can be denoted as

$$E_{OPEA}((\text{int})a \parallel (\text{int})b \parallel (\text{int})c \parallel (\text{int})d)$$

Here, each plaintext character is concatenated after being transformed into a 3-digit decimal ASCII code, where ‖ indicates a string concatenation. The ciphertext expansion in this rule will be limited when the plaintext is small. The order-preservation and the additivity are preserved for the field as well.

3) Field in equality, range and aggregation query.

For the fields in the equality, range and aggregation queries, the plaintext string can also be transformed by minimum matching unit into ASCII code and then encrypted in sequence. Besides, encrypted characters are concatenated together after being padded to a fixed length, so the encrypted string can be denoted as

$$E_{OPEA}((\text{int})a) \parallel E_{OPEA}((\text{int})b) \parallel E_{OPEA}((\text{int})c) \parallel E_{OPEA}((\text{int})d)$$

Specifically speaking, we assume that each ASCII character is encrypted to a $N$-digit integer. And a zero padding is utilized in the most left of the insufficient characters. The ciphertext expansion in this rule will be limited when the plaintext is large. Only the order-preservation is preserved for the field, but not the additivity.

The third rule is also suitable for encrypting long integers. In order to avoid the run out of bounds exception in ciphertext partition that caused by overlarge plaintext domain, a long integer can be treated as a string. It is divided into multiple segments, encrypted individually, and concatenated orderly. Other data types might be implemented by appropriate type conversion and rule extension.

The data storage type in encrypted databases is bigint and varchar. No encoding is adopted here in any form, which aims at preserving the data manipulability. When a database owner decrypts the OPEA ciphertext, she first splits the encrypted string using the standard delimiter or the fixed length. After being decrypted individually, the segments are concatenated together to get the original plaintext.

In addition, similar to traditional plaintext databases, outsourced encrypted databases also face the issue on incompatible table size. Taking SQL Server as an example, the cloud regulates that there are at most 1024 columns in a table and at most 8060 bytes in a row. Since the length of ciphertexts is influenced by the $T$ value, T-DB breaks the above restriction faster than plaintext databases. The solution lies in that a plaintext column is split into multiple columns and stored into a new table. The data item whose length is not enough in splitting is padded with zero. The database owner then encrypts them separately, which could also improve somewhat the outsourcing security.

## 7.4 Related Work

In 2002, Hacigümüs et al. [5] first presented the concept of DBaaS. From then on, under external and internal attacks, the issues on the security and privacy of outsourced data were identified as a vital research direction. A number of symmetric and asymmetric encryption techniques have been widely used in protecting the confidentiality and integrity of outsourced databases [46, 47, 48], but most of them end in the impracticability of data manipulations over ciphertexts. Current representative researches mainly include two categories, i.e., single data manipulation schemes and multiple data manipulation schemes.

### *7.4.1 Single Data Manipulation*

From the perspective of supported query functionalities, the existing studies about single data manipulation in outsourced encrypted databases can be generally classified into four categories, which are shown as follows:

1) Equality query.

In the earliest research about security protection of cloud databases [5], an outsourced database is encrypted by a deterministic encryption scheme (e.g., RSA [49]), which supports accurate equality comparison. However, the deterministic encryption technique directly leaks the equality pattern between data duplicates. Similarly, only equality testing is available in the cryptosystem [50] that was studied by Lu and Tsudik, while JOIN operator and comparison operators are not naturally supported. Besides, it might reveal access structure or sensitive attribute values by brute force attacks if the CSP colludes with the attacker in this scheme.

More discussions are focused on executing Equi-Join query over encrypted data. In 2008, Li and Chen proposed a distributed privacy-preserving Equi-Join protocol [51]. One recent work [10] utilizes the bilinear pairing during query processing, causing high computational overhead. Related equality queries on ciphertexts [8, 9] also inevitably leak some information about access pattern to the CSP, which causes a series of security attacks [52, 53].

2) Range query.

The original form of range query in encrypted databases is the TOP-$k$ query [54, 55]. Vaidya et al. [55] developed privacy-preserving $k$-anonymity policies in which the data are vertically partitioned instead of being encrypted, but they cannot be extended to encrypted databases. After that, based on the research in [50], a solution proposed by Wang et al. [13] prevents the brute force attack from the point of view of the cloud database. A trusted proxy is explored to encrypt data tuples and to realize basic Equality Query, Range Query, INSERT, UPDATE or DELETE, but it weakens the protection of cloud users' private data items. Tian et al. [56] then adopted a set of non-colluding Database Service Providers (DSPs) according to secret sharing strategies. A range query request is transformed by the Data Requester (DR) into encrypted form. They also constructed an ordered privacy preserving index to accelerate query speed and raise query precision. This kind of distributed database outsourcing protocol relies on the assumption of non-collusion and always has large overhead in terms of computation, communication and storage.

Different from the above distributed secure multiparty computation, the first order preserving encoding function proposed by Agrawal et al. [11] in 2004 can be directly applied into the encrypted database. It performs various kinds of range queries (e.g., ORDER BY clause, GROUP BY clause, comparison operators) and related statistical aggregate functions (e.g., COUNT, MIN, MAX). mOPE scheme [12] also supports Range Query, INSERT, UPDATE and DELETE, but it has low efficiency in the order comparison. As for another order preserving encryption algorithm [57] with random noise, CREATE, INSERT and some basic SELECT statements are translated by the trusted proxy into a privacy-preserving form. The security is compromised because its secret parameters are reused many times, and because of its quasi-linear encryption structure. Besides, Reddy and Ramachandram designed a Randomized Order Preserving Encryption scheme (called ROPE [28]). Although ROPE further improves the insert, delete, update and query function in terms of equality and range in mOPE, the frequent communication with proxy reduces its query efficiency. MV-OPES [58, 59] is another randomized encryption scheme that maps a plaintext value into different multiple ciphertext values to prevent statistical attacks. This method allows some parts of privacy-preserving relational operations to be executed over an encrypted database, including selection, sorting and the projection based on inequality. But some operations like join, aggregation, grouping, set difference and set union need to eliminate false positives after decrypting the data, which brings additional workload for the client.

3) Aggregation query.

Researchers usually employ homomorphic encryption techniques to solve the problem on aggregation queries or algebraic operations in encrypted databases. Although the ciphertexts generated by fully homomorphic encryption [20] can be involved directly in arbitrary computation without decrypting them, this technique is impracticable and prohibitively expensive in computation. As a compromise, Partially Homomorphic Encryption (PHE) has been widely studied. Some examples include Paillier [60] (supports additive aggregation operation), RSA [49] (supports multiplicative aggregation operation), PEKS [61] (supports keyword search on encrypted strings), etc. Such a PHE scheme is just designed to process one particular type of operation, that is, a cryptosystem that supports additive aggregation does not apply to any ciphertext comparison. We fill the gap in this paper by proposing a cipher tool that preserves the order for ciphertexts after additive operations. Then range queries and aggregation queries can be achieved simultaneously in the encrypted database.

The distance-based search (e.g., *k*NN algorithm, similarity retrieval, etc.) is another important aggregation query. In 2009, Wong et al. [62] presented an asymmetric scalar-product-preserving encryption (ASPE) for *k*NN queries. As shown in [63], this scheme is not secure and is vulnerable to chosen-plaintext attacks. Recently, a secure *k*NN query processing technique [14] based on mOPE [12] and a secure *k*NN protocol [15] which is suitable for mobile devices have appeared continuously. The security about data confidentiality, user's query requests and data access patterns is well guaranteed. Besides, a modified privacy homomorphism was proposed by Hu et al. [16] to protect data privacy and query privacy. The scheme maps a set of data operations (i.e., addition, subtraction, multiplication and division) in the plaintext domain to another set of operations in the ciphertext domain. Recoding and scrambling functions are also used to calculate the distance of ciphertexts, and to ensure their monotonicity and security. However, this privacy homomorphism is only available in distance-based queries with *R*-tree index and it has high cost in computation and communication.

4) Fuzzy query.

In the outsourced encrypted database, fuzzy query usually includes keyword search and similarity retrieval, which are used to request some more complex queries. In special cases, the keyword search can be achieved by some extensions of relational operations in the database. Until now, lots of studies on single-keyword search [64] and multi-keyword search [17, 65] have been proposed for ciphertexts. But they often suffer from potential information leakage and collusion attacks. Furthermore, they consume large amounts of computing resources. TEES (Traffic and Energy saving Encrypted Search) [18] is an efficient similarity retrieval algorithm at the cloud. It improves the system security and saves the energy, meanwhile the network traffics during the information retrieval are also significantly reduced. The TF-IDF value (Term Frequency-Inverse Document Frequency) is mapped to different ciphertexts with a certain probability. Thus, the statistical information of the term frequency is approximately a uniform distribution. As for the searchable encryption with multiple data owners [66], its key distribution and management are becoming more and more complicated with the rapid increasing of users and the explosive growth of data. Fuzzy matching search has a tolerance of possible keyword misspelling, minor typos and format inconsistencies. The ciphertext search in [67] constructed a fuzzy keyword set by measuring keyword similarity semantics. Bing Wang et al. [68] presented a multi-keyword fuzzy query scheme for cloud outsourced data by exploiting locality-sensitive hashing technique.

Nevertheless, if the above searchable encryption techniques are applied directly to keyword-based queries in the encrypted database, the limitation is that wildcard characters cannot be matched and regular expressions cannot be parsed. Taking CryptDB [22] as an example, its keyword search algorithm SEARCH [41] is only available in the full-word search. A search pattern string like '%*αβγ*%' can be correctly queried (where *αβγ* is a search keyword), while other forms of wildcard-based queries are ignored completely. So CryptDB does not support fuzzy queries in the genuine sense. To overcome this limitation, Dongsheng Wang et al. [19] proposed GPSE (Generalized Pattern-matching String-search on Encrypted data) for cloud encrypted data. GPSE allows the database users to request a query using generalized wildcard-based pattern strings (similar to the LIKE mode in SQL query). Its security has been demonstrated under known-plaintext model and high search accuracy has been achieved.

### *7.4.2 Multiple Data Manipulation*

At present, several integrated solutions for outsourced encrypted databases have been proposed. They attempt to provide multiple data manipulations in one-time.

Hakan Hacigümüs et al. [25] presented a query framework in 2002 to protect the data security in DBaaS for the first time. OPE and probabilistic encryption are adopted to support basic relational operations in equality, range and aggregation queries. It is fairly good in efficiency but relatively poor in precision. A relevant research [59] realized the same functionality simply by OPE, but a multitude of false positives still exist in its query results.

The latest SDB system [26, 69] is a series of secure multiparty computation protocols based on asymmetric secret sharing. Data interoperability is achieved in a wide range of queries by the interaction between a single DSP and the client. Even so, since its data model is always column-based, efficient comparison is hardly carried out on the same data column, that is, some statistical functions like MIN, MAX and TOP-*k* have not been fully performed. Furthermore, this system cannot encrypt character data, so fuzzy search is not supported and the PEKS cryptosystem [61] needs to be introduced.

In 2011, the best-known solution called CryptDB [22] was put forward. With the help of UDFs, a trusted proxy takes charge of translating a majority of query statements into executions on the encrypted database. The query result is roughly accurate. Though multiple ciphertexts are stored by the onion encryption, it can still be seen that CryptDB is too complicated for the encryption in the outsourced database, especially for large-scale aggregation operations. In addition, Wu et al. [70] designed a solution for the privacy-preserving queries over encrypted character data. By means of the format-preserving encryption technique, Li et al. [23] presented a lightweight database encryption mechanism L-EncDB. Similarly, specialized encryption algorithms (e.g., AES, OPE and FQE) are used respectively to interpret certain kinds of SQL statements. In most cases, exact query results can be obtained in this mechanism. Other studies [24, 71] developed from the above integrated queries are devoted to making plans for optimization strategies or scheduling for complex analytical query statements. Their target is to reduce the cost of query on ciphertexts and to improve the overall query efficiency.

Through the above summary, there exists a problem about the impracticability of ciphertext manipulations which should not be ignored in most of the existing solutions. Therefore, it is necessary for this paper to propose an almost fully functional, secure, efficient and friendly outsourcing scheme for the encrypted cloud databases.

## 7.5 Limitation and Future Work

T-DB attempts to find a trade-off between performance and security. On one hand, OPEA perfects the functionality of ciphertext query. For example, the results of aggregation query can be compared directly, thereby improving overall query performance. On the other hand, OPEA with IND-AOCPA achieves its optimal security under the same functionalities. As for the above case of aggregation-comparison, only the ciphertext range of the arithmetic sum in the sum calculation protocol is leaked to the CSP, which avoids possible leakage of the plaintext distance caused by self-increase operation.

The database owner performs the post-processing stage after receiving query results returned by the cloud. Decryption or secure two-party computation protocol is used to acquire the final query results for the database user. T-DB is limited by the arithmetical expression, especially the multiplication among attribute columns, and this is the main performance bottleneck of the post-processing stage. For example, in contrast to homomorphic techniques, the protocol in T-DB needs the involvement of database owners to execute one round of data interaction except decryption of the returned encrypted results. As the data size increases, several communication expenses increase at the database owner side. In order to reduce the network traffic, it may be more practical for the database owner to perform these complex arithmetical computations after decryption. A new problem brought on by this method is how to divide the data manipulation into parts that can be executed on the cloud server, and parts that must be executed on the database owner [24]. It is valuable for further improving the overall performance of T-DB.

The experiments show that the total response time of an outsourced encrypted database is affected by the network throughput, transmission delay, query processing and the size of the result set, where the query efficiency is also influenced by query types and, particularly, scheduling strategies of SQL Azure. Our future work will focus on the performance improvement at the cloud RDBMS in the executions over encrypted databases. As illustrated in Table 6, a query in an encrypted database takes more than twice as long as in a plaintext database. The main reason is that since numerous UDFs are invoked in the encrypted query condition, the number of comparisons grows exponentially at the cloud server side. Fixing this issue is dependent on a targeted adjustment from industries, including a systematic overhaul of the index structure, the comparison operators, etc. Before these problems are addressed, one may obtain fast ciphertext query processing by a comprehensive promotion of resources and configuration of cloud servers.

What’s more, along with the increasing development of DBaaS techniques, introducing T-DB system to a verifiable auditing outsourcing (inspired by [72, 73]), to an application with multi-datasource (inspired by the scheme in [74]) and to a distance-based keyword similarity search (inspired by the work in [75]) will be investigated in the future.

## 7.6 Source Code Illustration

The source code [35] has been made available for replication purposes only.

The online supplementary file contains the source codes, test cases, UDF definitions and the license for T-DB. These files are structured as follows:

1) Subdirectory T-DB.
   - a. OPEA.cpp and OPEA.h: The implementation of OPEA algorithm.
   - b. SQLTranslator.cpp and SQLTranslator.h: The implementation of SQL-Translator.
   - c. PostProcessing.cpp and PostProcessing.h: Operations in pre-processing and post-processing stage.
   - d. Randoms.cpp and Randoms.h: A pseudo-random number generator.
   - e. Util.cpp and Util.h: Some utility functions.
   - f. OutsourcedDB.cpp: Some test interfaces.
   - g. PlainSQL.sql: The plaintext SQL statements before translation.
   - h. CryptedSQL.sql: The ciphertext SQL statements after translation.
2) Subdirectory Test: Some cases to test SQL-Translator.
3) Subdirectory UDF: The build definitions of T-DB UDFs.
4) LICENSE: The license of the source code for T-DB.
5) GPL-3: A copy of the GNU General Public License.
6) README: Brief introduction, code status, environmental requirements as well as illustrations of build steps and run steps.

## 8. CONCLUSIONS

In this paper, we proposed T-DB as an almost fully functional encrypted database system in clouds. The database is outsourced in the form of ciphertext and almost all types of database manipulations can be directly executed on the encrypted data (updates are supported as well). Neither the CSP nor the database users will be disturbed due to the encryption of the outsourced database. Owners can migrate their databases to an unmodified DBaaS platform smoothly and safely. The database structure and the access mode are preserved. T-DB realizes a database-owner-controlled reliable cloud storage pattern by encrypting the outsourced relational databases, and simultaneously achieves secrecy and utilization, instead of a compromise between them. Ciphertext values are queried with good precision and high efficiency. Our T-DB promotes cloud computing security best practice and may inspire widespread interest in the field of database security. It is expected to have a positive impact on database and information security academia and industry.

## 9. ACKNOWLEDGMENTS

X.W. and Y.Z. were supported by National Key R&D Program of China (2016YFB0800703), National Natural Science Foundation of China (61272481, 61572460), National Information Security Special Projects of the National Development and Reform Commission of China [(2012)1424], Open Project Program of the State Key Laboratory of Information Security (2017-ZD-01), and the China 111 Project (B16037). Q.W. was supported by National Natural Science Foundation of China (61672083, 61370190).

All relevant data needed to evaluate the conclusions in this paper as well as the source code of T-DB are available in this paper and/or its online supplementary materials [35]. Additional data related to this paper may be requested from the authors.

## 10. REFERENCES

[1] L. Wei, H. Zhu, Z. Cao, X. Dong, W. Jia, Y. Chen, and A. V. Vasilakos. Security and privacy for storage and computation in cloud computing. *Inf. Sci.*, 258:371-386, 2014.

[2] M. Ali, S. U. Khan, and A. V. Vasilakos. Security in cloud computing: Opportunities and challenges. *Inf. Sci.*, 305:357-383, 2015.

[3] International data corporation. https://www.idc.com/ [accessed 1 January 2017].

[4] Technology business research. http://www.tbri.com/ [accessed 1 January 2017].

[5] H. Hacigümüs, B. Iyer, and S. Mehrotra. Providing database as a service. In *IEEE ICDE*, pages 29-38, 2002.

[6] M. Abourezq and A. Idrissi. Database-as-a-service for big data: An overview. *Int. J. Adv. Comput. Sci. Appl.*, 7(1):157-177, 2016.

[7] Z. Zheng, J. Zhu, and M. R. Lyu. Service-generated big data and big data-as-a-service: An overview. In *BigData Congress*, pages 403-410, 2013.

[8] G. D. Crescenzo, D. Cook, A. McIntosh, and E. Panagos. Practical private information retrieval from a time-varying, multi-attribute, and multiple-occurrence database. In *DBSec*, pages 339-355, 2014.

[9] S. Jarecki, C. Jutla, H. Krawczyk, M. Rosu, and M. Steiner. Outsourced symmetric private information retrieval. In *ACM CCS*, pages 875-888, 2013.

[10] H. Pang and X. Ding. Privacy-preserving ad-hoc equi-join on outsourced data. *ACM Trans. Database Syst.*, 39(3):23, 2014.

[11] R. Agrawal, J. Kiernan, R. Srikant, and Y. Xu. Order preserving encryption for numeric data. In *ACM SIGMOD*, pages 563-574, 2004.

[12] R. A. Popa, F. H. Li, and N. Zeldovich. An ideal-security protocol for order preserving encoding. In *IEEE S&P*, pages 463-477, 2013.

[13] S. Wang, D. Agrawal, and A. E. Abbadi. A comprehensive framework for secure query processing on relational data in the cloud. In *SDM*, pages 52-69, 2011.

[14] S. Choi, G. Ghinita, H. Lim, and E. Bertino. Secure kNN query processing in untrusted cloud environments. *IEEE Trans. Knowl. Data Eng.*, 26(11):2818-2831, 2014.

[15] Y. Elmehdwi, B. K. Samanthula, and W. Jiang. Secure k-nearest neighbor query over encrypted data in outsourced environments. In *IEEE ICDE*, pages 664-675, 2014.

[16] H. Hu, J. Xu, C. Ren, and B. Choi. Processing private queries over untrusted data cloud through privacy homomorphism. In *IEEE ICDE*, pages 601-612, 2011.

[17] N. Cao, C. Wang, M. Li, K. Ren, and W. Lou. Privacy-preserving multi-keyword ranked search over encrypted cloud data. *IEEE Trans. Parallel Distrib. Syst.*, 25(1):222-233, 2014.

[18] J. Li, R. Ma, and H. Guan. TEES: An efficient search scheme over encrypted data on mobile cloud. *IEEE Trans. Cloud Comput.*, 5(1):126-139, 2015.

[19] D. Wang, X. Jia, C. Wang, K. Yang, S. Fu, and M. Xu. Generalized pattern matching string search on encrypted data in cloud systems. In *IEEE INFOCOM*, pages 2101-2109, 2015.

[20] C. Gentry. Fully homomorphic encryption using ideal lattices. In *ACM STOC*, pages 169-178, 2009.

[21] X. Cao, C. Moore, M. O'Neill, E. O'Sullivan, and N. Hanley. Optimised multiplication architectures for accelerating fully homomorphic encryption. *IEEE Trans.Comput.*, 65(9):2794-2806, 2016.

[22] R. A. Popa, C. M. S. Redfield, N. Zeldovich, and H. Balakrishnan. CryptDB: Protecting confidentiality with encrypted query processing. In *ACM SOSP*, pages 85-100, 2011.

[23] J. Li, Z. Liu, X. Chen, F. Xhafa, X. Tan, and D. S. Wong. L-EncDB: A lightweight framework for privacy-preserving data queries in cloud computing. *Knowl.-Based Syst.*, 79:18-26, 2015.

[24] S. Tu, M. F. Kaashoek, S. Madden, and N. Zeldovich. Processing analytical queries over encrypted data. In *VLDB*, pages 289-300, 2013.

[25] H. Hacigümüs, B. Iyer, C. Li, and S. Mehrotra. Executing SQL over encrypted data in the database-service-provider model. In *ACM SIGMOD*, pages 216-227, 2002.

[26] Z. He, W. K. Wong, B. Kao, D. W. L. Cheung, R. Li, S. M. Yiu, and E. Lo. SDB: A secure query processing system with data interoperability. In *VLDB*, pages 1876-1879, 2015.
[27] A. Boldyreva, N. Chenette, Y. Lee, and A. O'Neill. Order-preserving symmetric encryption. In *EUROCRYPT*, pages 224-241, 2009.
[28] K. S. Reddy and S. Ramachandram. A new randomized order preserving encryption scheme. *Int. J. Comput. Appl.*, 108(12):41-46, 2014.
[29] Z. Liu, X. Chen, J. Yang, C. Jia, and I. You. New order preserving encryption model for outsourced databases in cloud environments. *J. Netw. Comput. Appl.*, 59(C):198-207, 2016.
[30] M. Naveed, S. Kamara, and C. V. Wright. Inference attacks on property-preserving encrypted databases. In *ACM CCS*, pages 644-655, 2015.
[31] D. Boneh, K. Lewi, M. Raykova, A. Sahai, M. Zhandry, and J. Zimmerman. Semantically secure order-revealing encryption: Multi-input functional encryption without obfuscation. In *EUROCRYPT*, pages 563-594, 2015.
[32] K. Lewi and D. J. Wu. Order-revealing encryption: New constructions, applications, and lower bounds. In *ACM CCS*, pages 1167-1178, 2016.
[33] A. Silberschatz, H. F. Korth, and S. Sudarshan. *Database System Concepts*. McGraw-Hill, New York, ed. 6, 2010.
[34] Data manipulation language (DML) statements (Transact-SQL). https://technet.microsoft.com/en-us/library/ms177591(v=sql.90).aspx [accessed 1 March 2017].
[35] T-DB system. http://www.nipc.org.cn/tdbsources.aspx [accessed 15 August 2017].
[36] Access SQL: Basic concepts, vocabulary, and syntax. https://support.office.com/en-us/article/Access-SQL-basic-concepts-vocabulary-and-syntax-444D0303-CDE1-424E-9A74-E8DC3E460671 [accessed 1 August 2017].
[37] Writing SQL queries: Let's start with the basics. https://technet.microsoft.com/en-us/library/bb264565(v=sql.90).aspx [accessed 1 August 2017].
[38] OpenSSL cryptography and SSL/TLS toolkit. http://www.openssl.org/ [accessed 1 March 2017].
[39] Boost C++ libraries. http://www.boost.org/ [accessed 1 March 2017].
[40] TPC-H. http://www.tpc.org/tpch/default.asp [accessed 1 March 2017].
[41] D. X. Song, D. Wagner, and A. Perrig. Practical techniques for searches on encrypted data. In *IEEE S&P*, pages 44-55, 2000.
[42] Microsoft Azure. https://www.azure.cn/ [accessed 1 March 2017].
[43] SQL database migration wizard v3.15.6, v4.15.6 and v5.15.6. http://sqlazuremw.codeplex.com/ [accessed 1 March 2017].
[44] S. Hong, H. Kim, and J. Chang. An efficient key management scheme for user access control in outsourced databases. *World Wide Web*, 20(3):467-490, 2017.
[45] Q. Yaseen, Y. Jararweh, B. Panda, and Q. Althebyan. An insider threat aware access control for cloud relational databases. *Cluster Comput.*, 20(3):2669-2685, 2017.
[46] M. Zhou, Y. Mu, W. Susilo, J. Yan, and L. Dong. Privacy enhanced data outsourcing in the cloud. *J. Netw. Comput. Appl.*, 35(4):1367-1373, 2012.
[47] S. Yu, C. Wang, K. Ren, and W. Lou. Achieving secure, scalable, and fine-grained data access control in cloud computing. In *IEEE INFOCOM*, pages 534-542, 2010.
[48] D. Agrawal, A. E. Abbadi, and S. Wang. Secure and privacy-preserving database services in the cloud. In *IEEE ICDE*, pages 1268-1271, 2013.
[49] R. L. Rivest, A. Shamir, and L. M. Adleman. A method for obtaining digital signatures and public-key cryptosystems. *Commun. ACM*, 21(2):120-126, 1978.
[50] Y. Lu and G. Tsudik. Privacy-preserving cloud database querying. *J. Int. Serv. Inf. Secur.*, 1(4):5-25, 2011.
[51] Y. Li and M. Chen. Privacy preserving joins. In *IEEE ICDE*, pages 1352-1354, 2008.
[52] M. S. Islam, M. Kuzu, and M. Kantarcioglu. Access pattern disclosure on searchable encryption: Ramification, attack and mitigation. In *NDSS*, 2012.
[53] M. S. Islam, M. Kuzu, and M. Kantarcioglu. Inference attack against encrypted range queries on outsourced databases. In *CODASPY*, pages 235-246, 2014.
[54] L. Xiong, S. Chitti, and L. Liu. Preserving data privacy in outsourcing data aggregation services. *ACM Trans. Internet. Techn.*, 7(3):1-28, 2007.
[55] J. Vaidya and C. Clifton. Privacy-preserving top-k queries. In *IEEE ICDE*, pages 545-546, 2005.
[56] X. Tian, C. Sha, X. Wang, and A. Zhou. Privacy preserving query processing on secret share based data storage. In *DASFAA*, pages 108-122, 2011.
[57] D. Liu and S. Wang. Nonlinear order preserving index for encrypted database query in service cloud environments. *Concurr. Comp. Pract. E.*, 25(13):1967-1984, 2013.
[58] H. Kadhem, T. Amagasa, and H. Kitagawa. A secure and efficient order preserving encryption scheme for relational databases. In *KMIS*, pages 25-35, 2010.
[59] H. Kadhem, T. Amagasa, and H. Kitagawa. MV-OPES: Multivalued-order preserving encryption scheme: A novel scheme for encrypting integer value to many different values. *IEICE Trans. Inf. Syst.*, 93(9):2520-2533, 2010.
[60] P. Paillier. Public-key cryptosystems based on composite degree residuosity classes. In *EUROCRYPT*, pages 223-238, 1999.
[61] D. Boneh, G. D. Crescenzo, R. Ostrovsky, and G. Persiano. Public key encryption with keyword search. In *EUROCRYPT*, pages 506-522, 2004.
[62] W. K. Wong, D. W. Cheung, B. Kao, and N. Mamoulis. Secure kNN computation on encrypted databases. In *ACM SIGMOD*, pages 139-152, 2009.

[63] B. Yao, F. Li, and X. Xiao. Secure nearest neighbor revisited. In *IEEE ICDE*, pages 733-744, 2013.
[64] M. Kuzu, M. S. Islam, and M. Kantarcioglu. Efficient similarity search over encrypted data. In *IEEE ICD*E, pages 1156-1167, 2012.
[65] C. Örencik and E. Savas. An efficient privacy-preserving multi-keyword search over encrypted cloud data with ranking. *IEEE Trans. Parallel Distrib. Syst.*, 32(1):119-160, 2014.
[66] S. D. C. D. Vimercati, S. Foresti, S. Jajodia, S. Paraboschi, and P. Samarati. Encryption policies for regulating access to outsourced data. *ACM Trans. Database Syst.*, 35(2):1-46, 2010.
[67] J. Li, Q. Wang, C. Wang, N. Cao, K. Ren, and W. Lou. Fuzzy keyword search over encrypted data in cloud computing. In *IEEE INFOCOM*, pages 441-445, 2010.
[68] B. Wang, S. Yu, W. Lou, and Y. T. Hou. Privacy-preserving multi-keyword fuzzy search over encrypted data in the cloud. In *IEEE INFOCOM*, pages 2112-2120, 2014.
[69] W. K. Wong, B. Kao, D. W. L. Cheung, R. Li, and S. M. Yiu. Secure query processing with data interoperability in a cloud database environment. In *ACM SIGMOD*, pages 1395-1406, 2014.
[70] Z. Wu, G. Xu, Z. Yu, X. Yi, E. Chen, and Y. Zhang. Executing SQL queries over encrypted character strings in the database-as-service model. *Knowl.-Based Syst.*, 35:332-348, 2012.
[71] C. Curino, E. P. C. Jones, R. A. Popa, N. Malviya, E. Wu, S. Madden, H. Balakrishnan, and N. Zeldovich. Relational cloud: A database-as-a-service for the cloud. In *CIDR*, pages 235-240, 2011.
[72] J. Wang, X. Chen, X. Huang, I. You, and Y. Xiang. Verifiable auditing for outsourced database in cloud computing. *IEEE Trans.Comput.*, 64(11):3293-3303, 2015.
[73] X. Chen, J. Li, J. Weng, J. Ma, and W. Lou. Verifiable computation over large database with incremental updates. *IEEE Trans.Comput.*, 65(10):3184-3195, 2016.
[74] W. Song, B. Wang, Q. Wang, Z. Peng, and W. Lou. Tell me the truth: Practically public authentication for outsourced databases with multi-user modification. *Inf. Sci.*, 387:221-237, 2017.
[75] R. Xu, K. Morozov, Y. Yang, J. Zhou, and T. Takagi. Efficient outsourcing of secure k-nearest neighbour query over encrypted database. *Comput. Secur.*, 69:65-83, 2017.